\documentclass[conference,compsoc]{IEEEtran}
\IEEEoverridecommandlockouts
\usepackage{cite}
\usepackage{amsmath,amssymb,amsfonts}
\usepackage{algorithmic}
\usepackage{float}
\usepackage{graphicx}
\usepackage{array}
\usepackage{tabularx}
\usepackage{textcomp}
\usepackage[table,gray]{xcolor}
\usepackage{hyperref}
\usepackage{cleveref}
\crefname{table}{Table}{Tables} 
\hypersetup{
    hidelinks,         
}
\def\BibTeX{{\rm B\kern-.05em{\sc i\kern-.025em b}\kern-.08em
    T\kern-.1667em\lower.7ex\hbox{E}\kern-.125emX}}

\begin{document}

\title{R+R: Security Vulnerability Dataset Quality Is Critical}

\author{\IEEEauthorblockN{Anurag Swarnim Yadav}
\IEEEauthorblockA{\textit{Department Of Computer Science} \\
\textit{University Of Florida}\\
Gainesville, FL, USA \\
anuragswar.yadav@ufl.edu}
\and
\IEEEauthorblockN{Joseph N. Wilson}
\IEEEauthorblockA{\textit{Department Of Computer Science} \\
\textit{University Of Florida}\\
Gainesville, FL, USA \\
jnw@ufl.edu}
}

\maketitle
\begin{abstract}
Large Language Models (LLMs) are of great interest in vulnerability detection and repair.
The effectiveness of these models hinges on the quality of the datasets used for both training and evaluation.
Our investigation reveals that a number of studies featured in prominent software engineering conferences have employed datasets that are plagued by high duplication rates, questionable label accuracy, and incomplete samples.
Using these datasets for experimentation will yield incorrect results that are significantly different from actual expected behavior.
For example, the state-of-the-art VulRepair Model, which is reported to have 44\% accuracy, on average yielded 9\% accuracy when test-set duplicates were removed from its training set and 13\% accuracy when training-set duplicates were removed from its test set.
In an effort to tackle these data quality concerns, we have retrained models from several papers without duplicates and conducted an accuracy assessment of labels for the top ten most hazardous Common Weakness Enumerations (CWEs).
Our findings indicate that 56\% of the samples had incorrect labels and 44\% comprised incomplete samples---only 31\% were both accurate and complete.
Finally, we employ transfer learning using a large deduplicated bug-fix corpus to show that these models can exhibit better performance if given larger amounts of high-quality pre-training data, leading us to conclude that while previous studies have over-estimated performance due to poor dataset quality, this does not demonstrate that better performance is not possible.
\end{abstract}


\section{Introduction}\label{Introduction}
There are many recent publications presenting systems that use neural models to detect and repair security vulnerabilities in programs\cite{chen2022neural,fu2022vulrepair,zhang2023pre,huang2023empirical,fu2024vision,mastropaolo2024training}.
In the past few months a number of reports have surfaced concerning the quality of the datasets employed to train and test these systems in an effort to characterize their performance\cite{mastropaolo2024training,ding2024vulnerability}.
We join these other works by reporting here on the many data quality problems embodied in what we will call the \emph{VulRepair dataset}, first employed in the work of Fu et al.\cite{fu2022vulrepair}
This dataset is a combination of Big-Vul\cite{fan2020ac} and CVEfixes\cite{bhandari2021cvefixes}.

We first encountered the data quality problems in the VulRepair dataset while attempting to understand the performance of several systems that employed it.
At first, we took the results at face value.
However, when replicating their behavior and looking at the specific repairs being made by the systems, it became clear that they exhibited effects associated with data leakage--the model was able to make repairs to \emph{vulnerable} code that required it to have information available only in the \emph{repaired} code.

Once we identified the existence of numerous duplicate samples in the data collection, we realized there might be other data quality issues as well.
Some of the problems in BigVul and CVEfixes datasets are reported by Ding et al.\cite{ding2024vulnerability} in the context of vulnerability detection, which deals primarily with the issue of duplicate samples that lead to data leakage.


This paper, on the other hand, reports initial results of our comprehensive review of the quality of the VulRepair data set and its effect on expected model performance.

A number of papers have been written employing the VulRepair dataset. The first is the report describing VulRepair in November 2022\cite{fu2022vulrepair}, which introduced this dataset's use in fine-tuning a CodeT5-based system. 
This same dataset was employed by Zhang et al. in an August 2023 paper evaluating CodeT5, CodeBERT, GraphCodeBERT, UniXcoder, and CodeGPT models trained on the VulRepair dataset\cite{zhang2023pre}.
Huang et al. employed the VulRepair dataset in a September 2023 paper benchmarking security-related tasks with CodeT5, CodeBERT, GraphCodeBERT, UniXcoder, and PLBART\cite{huang2023empirical}.

Fu et al. reported further work leveraging VRepair's pre-training bug-fix dataset in conjunction with the VQM vulnerability fine-tuning dataset for program repair tasks.
That report asserts that VRepair\cite{chen2022neural} was initially pre-trained on a dataset having 23,607 C/C++ functions, however, this claim appears to be incorrect as VRepair was pre-trained on a bug-fix dataset that has 534,848 training samples and 10,000 validation samples (see Chen et al.\cite{chen2022neural} last sentence of Section 4.21).
We conducted a brief analysis of both the pre-training bug-fix dataset and VQM vulnerability fine-tuning dataset\cite{vqm2024} used in that paper. The pre-training dataset contains 21,246 training samples and 2,362 validation samples.
Our review revealed 18,622 duplicated entries in the training set and 782 duplicates in the validation set.
After removing those, 1,579 cross-set duplicates (in both train and validation) were identified, that is to say, \textit{all of the validation set samples} were present in the training set. 
Additionally, our analysis uncovered a substantial overlap between the bug-fix dataset and the VQM vulnerability fine-tuning dataset. Specifically, there were 511 matching entries in the test set, 243 in the validation set, and 1,747 in the training set of the vulnerability dataset that overlapped with the bug-fix dataset. 
The detailed analysis is discussed in Section \ref{Related-Work}.
These findings make it clear that retraining is necessary in order to understand the expected performance of the models reported in that work.


In this paper, we analyze the effect some of the data quality problems in the VulRepair dataset have had on skewing the expected performance of VulRepair, CodeBERT, GraphCodeBERT, and UniXcoder when applied to repair of security vulnerabilities.
In addition we consider the impact of pre-training models with the deduplicated VRepair bug-fix dataset, addressing the duplicate entries present in the original VRepair pre-training dataset as discussed in Section \ref{Pre-training}.

Our analysis incorporates the following steps:
\begin{enumerate}
    \item Replicate the results of previous papers using a more robust approach involving finding the mean performance of multiple models fine-tuned using different seeds.
    \item Determine the impact of duplicate samples on the performance of these models.
    \item Determine the performance impact of inconsistent labeling.
    \item Partially identify and analyze the inaccurate labeling.
    \item Partially identify and analyze the incomplete data samples.
    \item Determine the impact of pre-training with the deduplicated VRepair bug-fix dataset. 
\end{enumerate}

These data-quality-driven analyses are addressed by the following research questions:

\begin{itemize}
\item {\textbf{RQ1}}: Can we replicate the results of VulRepair, CodeBERT, GraphCodeBERT, and UniXcoder?

\item {\textbf{RQ2A}}: What is the performance of the models after removing all duplicate data from the VulRepair dataset (removing cross-set duplicates from the \emph{training} dataset)?

\item {\textbf{RQ2B}}: What is the performance of the models after removing all duplicates from the VulRepair dataset (removing cross-set duplicates from the \emph{test} dataset)?

\item {\textbf{RQ3A}}: What is the performance of the models after removing duplicates and inconsistently labeled samples from the VulRepair dataset (removing cross-set duplicates and inconsistent samples from the \emph{training} dataset)?

\item {\textbf{RQ3B}}: What is the performance of the models after removing duplicates and inconsistently labeled samples from the VulRepair dataset (removing cross-set duplicates and inconsistent samples from the \emph{test} dataset)?

\item{\textbf{RQ4A}}: What effect does removing duplicate data from VulRepair have on the repair of vulnerabilities included in the top ten CWEs.

\item {\textbf{RQ4B}}: How does identifying and analyzing inaccurate and incomplete samples of the top ten CWEs in the test set of the deduplicated VulRepair dataset change our understanding of its performance?


\item {\textbf{RQ5}}: What is the performance of the systems after pre-training with a deduplicated bug-fix dataset?

\end{itemize}

Scripts used to verify data duplication and to replicate experiments are available from \href{https://github.com/Anurag-Swarnim-Yadav/DatasetQuality}{https://github.com/Anurag-Swarnim-Yadav/DatasetQuality}.

\section{Dataset Quality}

In this section, we briefly review a few properties that must be satisfied by samples contained within a dataset in order to ensure that systems trained on it do not overestimate their performance on unseen data as a result of poor generalization.
A more complete discussion of this topic can be found in Croft et al.\cite{croft2023data} which deals with datasets for software vulnerability detection.
The data quality attributes discussed by Croft et al. are standardized in the ISO/IEC 25012 data quality framework\cite{iso/iec25012}.
They employ these attributes in an analysis of several datasets including Big-Vul.
Let's consider the data quality attributes relevant to our work.

\subsection{Uniqueness}
A datum is \emph{unique} if it is represented exactly once within the dataset.
Supervised learning systems typically employ datasets that are partitioned into one or more of training, validation, and test sets.
Duplicates within any of these subsets can lead to biased model performance.
Duplicates across these subsets lead to data leakage which can have the negative result of overestimating model performance.

\subsection{Consistency}
In the context of software vulnerability repair datasets, a datum is \emph{consistent} if it does not appear in the dataset with conflicting labels.
There are two possible reasons for a single code sample to be labeled in different ways in a data set:

\begin{enumerate}
    \item The labels employed do not partition the space of possible data, that is, they are not exclusive of each other.

    \item At most one of the labels for such a sample is the appropriate label and all other labels are incorrect.
\end{enumerate}

Inconsistent labels arising from either of these causes are inappropriate for machine learning systems intended to identify or repair security vulnerabilities.
If the system is to identify the unique class associated with a program vulnerability, it must be trained using samples with only a single label.
On the other hand, if the system is to repair such vulnerabilities, then inconsistent labels will make it more difficult to learn class-specific repairs. 

\subsection{Accuracy}

A datum in a dataset is \emph{accurate} if it embodies \emph{semantic label correctness}.
In the setting of a vulnerability dataset aimed at program repair, this means samples must correctly identify code as being vulnerable or correct.
If the dataset identifies the particular vulnerability exhibited by code, that vulnerability must be present in the sample.
If the dataset marks the vulnerable code, it must mark it correctly.
If a repaired sample is provided, the vulnerability must be correctly repaired.
And if the dataset marks the repaired code, it must mark it correctly.

Numerous samples in the VulRepair dataset are labeled with the wrong CWE tag.
This occurs for a variety of reasons.
In some cases, a change appears in a commit that addresses a CVE but the specific change is unrelated to that CVE.
In other cases the CWE labels from Big-Vul and CVEfixes do not agree as a result of a change to the NIST database during the time between the creation of the two data collections.

In some cases, correct code is labeled as being incorrect.
Other samples contain incomplete target patches.
In addition to such labeling problems, there are meta-problems associated with the sample labeling approach adopted by these datasets.
One such case is samples in which code must be moved in order to address a vulnerability.
Applying \texttt{<Startbug>} and \texttt{<EndBug>} labels to such incorrectly placed code is conflated with the use of these tags to identify code that must be changed in-place.
Perhaps we need an alternate mechanism to identify repairs that involve code motion rather than modification.

\subsection{Completeness}
A vulnerability repair dataset sample is considered to be \emph{complete} if the sample itself contains all information necessary to determine that the identified vulnerability exists and is correctly repaired.

\subsection{Discussion}
\label{Dataset:Discussion}

When Fu et al. published their work on VulRepair, their dataset was assigned two different quality certifications by ACM, namely, \emph{Artifacts Available} and \emph{Artifacts Evaluated - Reusable}\cite{acm-badges}.
The second of these badges is to be assigned only to artifacts that have successfully completed ``an independent audit.''
The nature of the audit is not reported, but clearly that audit did not effectively assess the data quality attributes discussed above.

Construction of datasets containing the numbers of samples necessary to effectively fine-tune code language models to deal with security vulnerability repair is a difficult task.
Attempts to construct such data sets in an automated way have been fraught with data quality problems.
We discuss our recommendations concerning construction and maintenance of such datasets in Section \ref{Conclusion}.
\section{Methodology}\label{Methodology}

\subsection{Dataset}

\subsubsection{Types of Duplicates}
A training dataset can have two types of duplicates: in-set duplicates and cross-set duplicates. Both categories introduce bias in some form.
An in-set duplicate is a sample that appears multiple times within one of either the training, validation, or test set.
Test-set duplicates can distort performance estimates, either positively or negatively, based on whether the duplicate is correctly or incorrectly repaired. 
Validation set duplicates can cause training to stop prematurely or continue longer than necessary, depending on whether the duplicates are correctly repaired.
Training set duplicates introduce a slight bias toward favoring correct results on duplicated entries over non-duplicated ones.
Cross-set duplicates are identical samples that are present in two or more of the training, validation, and test sets.
Cross-set duplicates introduce data leakage if samples from the training or validation sets are present in the test set, as the model has already encountered those samples during training.
As a result, the model can memorize responses to these specific instances, leading to a high evaluation score, overfitting, and a compromised ability to generalize.
To ensure our analysis was free of duplicates, we implemented a two-step process.
First, we eliminated in-set duplicates within each of the three datasets (training, validation, and test sets).
This step removed repeated entries within each set.
Next, we addressed cross-set duplicates by removing any instances that appeared in multiple sets.
This process prevents data leakage, ensuring that no data from the training set appears in the validation or test sets, thereby improving the validity of our model evaluation.

\subsubsection{Method for Identifying Duplicates}
The simplest way to identify  duplicates is to perform an exact string match between the samples or compare string hashes.
We have implemented deduplication by applying the Pandas drop\_duplicates() method to the input data.
Pandas drop\_duplicates() performs an element-wise comparison of the data in the DataFrame to identify exact duplicates.
We kept the first occurrence of a duplicate in each case.
This addresses the uniqueness attribute discussed in Section \ref{Uniqueness} but the VulRepair dataset also suffers from consistency problems discussed in Section \ref{Consistency}.
To ensure consistency, we first remove the CWE pre-tag from all input samples and apply the drop\_duplicates() method to help identify more duplicate samples.
After removing the duplicates, we restore CWE pre-tags to the appropriate input samples.

\subsubsection{Pre-training} \label{Pre-training}
Chen et al.\cite{chen2022neural} introduced VRepair, a neural network model designed for vulnerability repair using a vanilla transformer architecture.
The model undergoes pre-training with a generic bug-fix corpus followed by fine-tuning on a specific vulnerability dataset.
In the bug-fix corpus, each input buggy function is prefixed with a \texttt{CWE-\emph{XXX}} tag, which denotes the type of CWE category, followed by the buggy function itself.
The sections of the code in the buggy function that contain bugs are demarcated with \texttt{<S2SV\_StartBug>} and \texttt{<S2SV\_EndBug>} tags.
The \texttt{<S2SV\_StartBug>} tag indicates the beginning of the buggy snippet, while \texttt{<S2SV\_EndBug>} marks its end.
As the pre-training corpus is not tailored to specific bug types, all buggy functions are tagged with \texttt{CWE-000}.
The corresponding corrections in the patched versions of these snippets are marked with \texttt{<ModStart>}, indicating the start of the corrected code, and end with \texttt{<ModEnd>}, marking the conclusion of the modification that repairs the vulnerability.
Table \ref{Methodology:bug-fix-dataset} presents the dataset overview, featuring 534,858 training samples and 10,000 validation samples.
Our analysis identified 6,192 \emph{in-set} duplicates in the training set and 4 in the validation set, as well as 247 \emph{cross-set} duplicates.
The presence of such duplicates points out the need for a thorough dataset cleaning before using it for pre-training.
\begin{table}[h]
\centering
  \caption{Bug-fix Dataset Analysis}
  \label{Methodology:bug-fix-dataset}
  \begin{tabular}{|l|r|r|}
    \hline
    Samples &Train &Validation\\
    \hline
    Total Samples (TS)                                  &534,858  &10,000\\
    \hline
    In-Set Duplicates (IS Dup)                          &6,192  &4\\
    \hline
    Samples Left (SL = TS - IS Dup)                     &528,666  &9,996\\
    \hline
    {Cross-Set Duplicates (CS Dup)}  &\multicolumn{2}{c|}{247}\\
  \hline
\end{tabular}
\end{table}

\subsubsection{Fine-tuning}
For all of our fine-tuning experiments, we use the VulRepair \cite{fu2022vulrepair} dataset provided on Hugging Face\cite{VulRepair-HuggingFace}.
The dataset was created by combining two existing vulnerability datasets, Big-Vul\cite{fan2020ac} and CVEfixes\cite{bhandari2021cvefixes}.
Both datasets were gathered by crawling CVE databases and extracting information such as CWE ID, CVE ID, and git commit link related to the vulnerabilities, with the Big-Vul dataset encompassing vulnerabilities from 2002 to 2019, and CVEfixes covering vulnerabilities from 1999 to 2021.
The VulRepair dataset consists of a train and test set.
As with VRepair, each input in both the train and validation set is prefixed with a \texttt{CWE-\emph{XXX}} tag, which denotes the CWE category of the vulnerability, followed by the vulnerable function.
The training set consists of samples that span 93 unique CWEs and the test set spans 74 unique CWEs.
The sections of the vulnerable code that contain vulnerabilities are demarcated with \texttt{<S2SV\_StartBug>} and \texttt{<S2SV\_EndBug>} tags and the corrected code is marked with \texttt{<ModStart>} and \texttt{<ModEnd>} as per Chen et al. described above.
Table \ref{Methodology:VulRepair-dataset-Unique} presents an overview of the dataset.
The dataset has 6,776 training samples and 1,707 test samples.

Our examination of the uniqueness of the VulRepair dataset samples revealed the presence of numerous in-set and cross-set duplicates.
Overall, we identified 1593 duplicates in the training set, 91 duplicates in the test set, and 796 cross-set duplicates.
The prevalence of duplicates can largely be attributed to the overlap of samples resulting from merging the Big-Vul and CVEfixes which contain many of the same samples with conflicting labels.
\begin{table}[h]
\centering
  \caption{VulRepair Dataset Analysis-Uniqueness}
  \label{Methodology:VulRepair-dataset-Unique}
  \begin{tabular}{|l|r|r|}
    \hline
    Samples &Train &Test\\
    \hline
    Total Samples (TS)                                  &6,776  &1,706\\
    \hline
    In-Set Duplicates (IS Dup)                          &1,593  &91\\
    \hline
    Samples Left (SL = TS - IS Dup)                     &5,183  &1,615\\
    \hline
    {Cross-Set Duplicates (CS Dup)}  & \multicolumn{2}{c|}{796}\\
  \hline
\end{tabular}
\end{table}

We performed another analysis addressing labeling consistency in the VulRepair dataset in which we removed the \texttt{CWE-\emph{XXX} tags} from the input sequences in both training and test sets, to reveal numerous samples we refer to as \emph{CWE-duplicates}.
Table \ref{Methodology:VulRepair-dataset-Consistency} presents our analysis. 
\begin{table}
\centering
  \caption{VulRepair Dataset Analysis-Consistency}
  \label{Methodology:VulRepair-dataset-Consistency}
  \begin{tabular}{|l|r|r|}
    \hline
    Samples &Train &Test\\
    \hline
    Total Samples (TS)                                  &6,776  &1,706\\
    \hline
    In-Set Duplicates (IS Dup)                          &1,858  &111\\
    \hline
    Samples Left (SL = TS - IS Dup)                     &4,918  &1,595\\ 
    \hline
    {Cross-Set Duplicates (CF Dup)}  &\multicolumn{2}{c|}{923}\\
  \hline
\end{tabular}
\end{table}

Compared to Table \ref{Methodology:VulRepair-dataset-Unique}, we identified an additional 265 (1858-1593) in-set duplicates in the training set and  20 (111-91) in the test set, as well as  127 (923-796) more cross-set duplicates.
This analysis underscores the necessity of addressing both duplicate and mislabeled entries to ensure the effectiveness of machine learning models in vulnerability repair tasks.

\subsection{Models}
\subsubsection{VulRepair/CodeT5:} CodeT5\cite{wang2021codet5} is a pre-trained encoder-decoder model that extends the T5 architecture by incorporating token-type information tailored to coding languages.
It is pre-trained with the CodeSearchNet dataset, which has both programming language-only as well as programming and natural language data. Further refinement of the model is conducted through fine-tuning tasks such as code defect detection and summarization.

\subsubsection{CodeBERT:} CodeBERT\cite{feng2020codebert} is a bimodal, pre-trained model adept at processing both natural language (NL) and various programming languages (PL).
It incorporates two key components: standard masked language modeling (MLM) and replaced token detection.
In MLM, a certain percentage of the input tokens are randomly masked, and the model is trained to predict the original tokens based solely on the context provided by the remaining unmasked tokens.
The replaced token detection approach extends this by replacing some tokens with other plausible alternatives, challenging the model not only to predict the original tokens but also to identify which tokens have been altered.
Implementation of these components enhances CodeBERT and boosts the model's efficiency and effectiveness for handling unimodal and bimodal tasks.
CodeBERT is trained using data from GitHub repositories across six programming languages, including function-level natural language documentation alongside the corresponding code.

\subsubsection{GraphCodeBERT:} GraphCodeBERT\cite{guo2020graphcodebert} is a pre-trained model focusing on semantic-level structures over syntactic-level structures in a programming language.
In addition to implementing masked language modeling (MLM), it incorporates structure-aware pre-training tasks.
These tasks enable the model to comprehend and anticipate data flows and dependencies within the code, aligning the learned representations of the raw source code with its corresponding structural (data flow) representations.
This enhances its capacity to interpret and generate code, taking into account both its textual and structural elements.

\subsubsection{UniXcoder:} UniXcoder\cite{guo2022unixcoder} is a cross-modal pre-trained model designed specifically for programming languages, aimed at enhancing both code-related understanding and generation tasks.
The model undergoes pre-training using three distinct language modeling tasks: masked language modeling, unidirectional language modeling, and a denoising objective.
To improve code representation, UniXcoder integrates multiple modalities, including Abstract Syntax Trees (AST) and code comments.
It employs mask attention matrices combined with prefix adapters to guide model behavior and utilizes a one-to-one mapping method to convert the tree-structured AST into sequential formats.
This multifaceted approach allows UniXcoder to capture a deeper understanding of code semantics and structure.

\subsection{Evaluation Metrics}
We use perfect prediction/exact-match as implemented by Fu and Zhang et al.\cite{fu2022vulrepair,zhang2023pre} to evaluate the performance of these models.

\subsection{Experiment Setup}
\subsubsection{Pre-training hyperparameter settings for bug-fix}\label{Pre-training-hyperparameters}
For the pre-training of our dataset, we employed the default parameters consistent across the prior studies.
We used the Adam optimizer with a learning rate of 2e-5, and set the input and output sequence lengths to 512 and 256, respectively.
The training is designed to run for 75 epochs; however, training will stop early if the evaluation loss does not change over three consecutive epochs.
This early stopping mechanism helps prevent overfitting.
The batch size for both training and evaluation is set at 128.
During inference, we experimented with beam sizes 1,3,5 and 50 to explore the impact of beam size on model performance, which is detailed in Section \ref{TransferLearning}.

\subsubsection{Fine-tuning hyperparameter settings for VulRepair}\label{Fine-tuning-hyperparameters}
We adhered to the standard training practices established in previous research for the fine-tuning phase. The Adam optimizer was used with a learning rate of 2e-5 and the input and output sequence lengths were maintained at 512 and 256, respectively. The training duration was set for 75 epochs. A fixed beam size of 50 was used during inference, following the practices of prior studies. For transfer learning, we assessed the model with a collection of beam sizes ranging from 1 to 50 to determine their effects. These findings are discussed in Section \ref{TransferLearning}.

\subsubsection{Hardware parameters}
All training and evaluation was performed on NVIDIA A100-SXM4-80GB GPUs.

\subsection{Mean Performance}

In an attempt to provide robust performance evaluations, each result is reported as the mean performance of six networks trained using different random seeds.
We have noted significant performance differences in which perfect prediction rate can vary by more than one and one half times from one seed to another.

\subsection{Data Attributes Analysis}\label{Data-Attributes-Analysis}
Determining the accuracy and completeness attributes for each sample was performed manually, as no definitive oracle exists to guarantee the presence or absence of vulnerabilities in a specific piece of code.
Our analysis is confined to individual functions rather than the entire code base, which means our measures of accuracy and completeness are restricted to the function level.
Our primary objective is to confirm that the identified buggy function associated with a particular CWE vulnerability clearly contains the vulnerability and is accurately classified under the correct CWE.
After verifying the accuracy, we evaluate the available code changes.

 We implemented the following subsequent steps to ascertain the functional relevance of the identified vulnerable sample in relation to the assigned CWE label.
 \begin{enumerate}
\item For each instance in the VulRepair dataset, extract the CVE-ID and code commit that addresses the vulnerability.
The CVE-ID is crucial for cross-validation with the NVD database, which serves as the foundation for the VulRepair database.
\item Review pertinent details related to the vulnerable sample, such as a description of the vulnerability and applied fix as documented in the commit message.
\item Analyze the code changes in a code commit as well as the entire function to understand the relevance of the assigned CWE label.
\item Examine the fixed code to ensure it contains comprehensive information necessary for the relevant code changes.
\item Scrutinize the data attributes of the vulnerable sample.
\item If the data attributes could not be examined, we classified such samples as \textit{unverifiable samples}. This classification indicates that either the vulnerable sample or the repaired code lacked the necessary information to determine the data attributes.
 \end{enumerate}

\section{Results}\label{Result}

We first replicate the results of the respective previous works to get a baseline performance estimate on the VulRepair dataset.
We follow that by eliminating defective samples in the VulRepair set that fail to satisfy the data quality attributes of uniqueness and consistency, and then we analyze samples for accuracy and completeness.
We follow that by investigating the impact of transfer learning on model performance.

\subsection{Replication}

This experiment aims to replicate the results of code repair methods whose performance on the VulRepair dataset has been reported.

\textbf{RQ1: Can we replicate the result of VulRepair, CodeBERT, GraphCodeBERT, and UniXcoder?}

\begin{table}[h]
\centering
  \caption{VulRepair Hugging Face Dataset Overview}
  \label{Methodology:RQ1-dataset}
  \begin{tabular}{|l|r|l|}
    \hline
    Dataset&Number of Samples&Comments\\
     \hline
    Train & 6,776    & Contains IS and CS duplicates\\
     \hline
    Test  & 1,706    & Contains IS and CS duplicates\\
   \hline
\end{tabular}
\end{table}

Table \ref{Methodology:RQ1-dataset} shows the number of samples in the VulRepair dataset.

\textbf{Result:} The detailed results of this experiment are shown in Table \ref{Appendices:RQ1} and recapped in Table \ref{Methodology:RQ1-dataset-ModelPerformance} which shows reported perfect prediction (\emph{PP}) scores (see citations in the table) and our replicated scores.
The mean replicated performance for VulRepair is somewhat lower than published reports while CodeBERT, GraphCodeBERT, and UniXcoder are slightly higher.

\begin{table}[h]
\centering
  \caption{Model Performance on RQ1}
  \label{Methodology:RQ1-dataset-ModelPerformance}
  \begin{tabular}{|l|r|r|r|}
    \hline
    Models          & PP Reported    & PP Replicated & Change\\
    \hline
    VulRepair/CodeT5       & 44\%\cite{fu2022vulrepair}     & 40.42\% & -3.58\%\\
    &44.96\%\cite{zhang2023pre}  & & -4.54\%\\
    \hline
    CodeBERT        & 31\%\cite{fu2022vulrepair}      & 33.20\% & +2.20\%\\
    & 32.94\%\cite{zhang2023pre} && +0.74\%\\
    \hline
    GraphCodeBERT   & 37.98\%\cite{zhang2023pre}      & 38.51\%\ & +0.53\%\\
    \hline
    UniXcoder       & 40.62\%\cite{zhang2023pre}      & 40.96\% & +0.34\%\\
  \hline
  \multicolumn{4}{l}{Note: Huang et al.\cite{huang2023empirical} results are NOT reported as they used}\\
  \multicolumn{4}{l}{PPL and BLUE metrics instead.}\\
\end{tabular}
\end{table}






\subsection{Uniqueness}\label{Uniqueness}

In these experiments, we consider the effect of remediating VulRepair dataset uniqueness defects on the performance of the models fine-tuned with that dataset.

 In order to make minimal changes to the semantic content of the dataset, we preserve all unique data in the training and test sets---but cross-set duplicates must be removed from one set or the other.
This means the proportion of samples in the training and test sets will vary from experiment to experiment and will not match that of the original dataset.
To understand the resulting bias in our estimates of performance, we run the models in two different ways to determine the impact of first removing the cross-set duplicates from the \emph{training} set and then removing them from the \emph{test} set.
Our hypothesis was that removing them from the test set will yield poorer performance because test samples that were subject to data leakage would be less likely to be correctly repaired.
Furthermore we would expect that removing such samples from the training set will yield even poorer performance because of the significant reduction in training samples. 
We preserve all unique elements in both the training and test sets.
The number of samples in these modified train/test sets are reduced by this process, however, the amount of information in this modified set is unchanged from the original set because every unique sample is included.

\textbf{RQ2A: 
What is the performance of the models after removing all duplicates data from the VulRepair dataset (removing cross-set duplicates from the \emph{training} dataset)?}

For this research question, we first removed in-set duplicates and then removed cross-set duplicates from the training dataset to avoid data leakage issues and keep the test set as large as possible. Table \ref{Methodology:RQ2-removed-crossfile-from-train} presents our analysis.

\begin{table}[h]
\centering
 \caption{RQ2A Dataset Composition (removing cross-set duplicates from the training set). See Table \ref{Methodology:VulRepair-dataset-Unique} for SL and CS DUP.}
 \label{Methodology:RQ2-removed-crossfile-from-train}
 \begin{tabular}{|l|r|r|}
   \hline
   Samples &Train &Test\\
   \hline
   Unique Samples (US = SL - CS DUP)                    &4,387   &1,615\\
 \hline
\end{tabular}
\end{table}

\begin{table}[h]
\centering
  \caption{Model Performance on RQ2A}
  \label{Methodology:RQ2A-dataset-ModelPerformance}
  \begin{tabular}{|l|r|r|}
    \hline
    Model           & PP RQ2A & \% PP Replicated\\
    \hline
    VulRepair       & 8.91\% & 22.0\% (8.91/40.42)\\
    \hline
    CodeBERT        & 5.58\% & 16.8\% (5.58/33.20)\\
    \hline
    GraphCodeBERT   & 5.31\% & 13.7\% (5.31/38.51)\\
    \hline
    UniXcoder       & 4.82\% & 11.8\% (4.82/40.96)\\
  \hline
\end{tabular}
\end{table}

\textbf{Result:} Detailed results of this experiment are shown in Table \ref{Appendices:RQ2A} and recapped in Table \ref{Methodology:RQ2A-dataset-ModelPerformance} where column \emph{PP RQ2A} shows perfect prediction scores on running on RQ2A dataset and \emph{\% PP Replicated} shows the fraction of perfect prediction in our replicated results from the VulRepair dataset shown in Table \ref{Methodology:RQ1-dataset-ModelPerformance}.
All models suffered extreme performance degradation when duplicates were removed from the dataset, providing perfect prediction rates ranging from 12 to 22\% of the rates they achieved when trained with a set that exhibited data leakage.  

\textbf{RQ2B: What is the performance of the models after removing all duplicates from the VulRepair dataset (removing cross-set duplicates from the \emph{test} dataset)?}

In this research question, we removed cross-set duplicates from the test dataset, keeping the train set as large as possible. Table \ref{Methodology:RQ2B-removed-crossfile-from-test} presents our analysis.

\begin{table}[h]
\centering
 \caption{RQ2B Dataset composition (removing cross-set duplicates from the test set). See Table \ref{Methodology:VulRepair-dataset-Unique} for SL and CS DUP.}
 \label{Methodology:RQ2B-removed-crossfile-from-test}
 \begin{tabular}{|l|r|r|}
   \hline
   Samples &Train &Test\\
   \hline
   Unique Samples (US = SL - CS DUP)                    &5,183   &819\\
 \hline
\end{tabular}
\end{table}

\begin{table}[h]
\centering
  \caption{Model Performance on RQ2B}
  \label{Methodology:RQ2Bdataset-ModelPerformance}
  \begin{tabular}{|l|r|r|}
    \hline
    Models          & PP RQ2B & \% PP Replicated\\
    \hline
    VulRepair       & 13.17\% & 33\% (13.17/40.42)\\
    \hline
    CodeBERT        & 8.83\% &  27\% \enspace(8.83/33.20)\\
    \hline
    GraphCodeBERT   & 9.22\% & 24\% \enspace(9.22/38.51)\\
    \hline
    UniXcoder       & 9.10\% & 22\% \enspace(9.10/40.96)\\
  \hline
\end{tabular}
\end{table}

\textbf{Result:}
As predicted, performance of all models was much better in RQ2B than  RQ2A. Detailed results are shown in Table \ref{Appendices:RQ2B} and recapped in Table \ref{Methodology:RQ2Bdataset-ModelPerformance}.
The increase in training samples from 4,387 to 5,183 provided perfect prediction increases of about 50\% over RQ2A in all cases, but still much lower (a third or less) than the performance when trained with the VulRepair dataset.

\subsection{Consistency}\label{Consistency}
In this section, we consider the impact that inconsistent labeling has on these models using the same two approaches as before.
Recall that inconsistent labels are associated with samples that have the identical buggy and repaired code, but different CWE labels.
Again, we remove these from the training set in one experiment and the test set in another.

\textbf{RQ3A: What is the performance of the models after removing duplicates and inconsistently labeled samples from the VulRepair dataset (removing cross-set duplicates and inconsistent samples from the \emph{training} dataset)?}

Table \ref{Methodology:RQ3A--removed-samples-with-multilple-CWEs} presents our analysis. See Table \ref{Methodology:VulRepair-dataset-Consistency} for SL and CS DUP.

\begin{table}[h]
\centering
  \caption{RQ3A Dataset composition (removing cross-set duplicates and inconsistent samples from the training set)}
  \label{Methodology:RQ3A--removed-samples-with-multilple-CWEs}
  \begin{tabular}{|l|r|r|}
    \hline
    Samples &Train &Test\\
    \hline
    Unique Samples (US = SL - CS DUP)                    &3,995  &1,595\\
  \hline
\end{tabular}
\end{table}

\begin{table}[h]
\centering
\caption{Model Performance on RQ3A}
  \label{Methodology:RQ3Adataset-ModelPerformance}
  \begin{tabular}{|l|r|r|}
    \hline
    Models          & PP RQ3A & \% PP Replicated\\ 
    \hline
    VulRepair       & 7.14\%  & 17.7\% (7.14/40.42)\\
    \hline
    CodeBERT        & 3.59\% & 10.8\% (3.59/33.20)\\
    \hline
    GraphCodeBERT   & 3.75\% & 9.7\% (3.75/38.51)\\
    \hline
    UniXcoder       & 4.11\% & 10.0\% (4.11/40.96)\\
  \hline
\end{tabular}
\end{table}

\textbf{Result:}
Detailed results of this experiment are shown in Table \ref{Appendices:RQ3A}. Table \ref{Methodology:RQ3Adataset-ModelPerformance} shows that all models suffered further performance degradation from the results shown in RQ2A.
This suggests that data leakage from the sample code occurs even when the CWE tag is incorrect.

\textbf{RQ3B: What is the performance of the models after removing duplicates and inconsistently labeled samples from the VulRepair dataset (removing cross-set duplicates and inconsistent samples from the \emph{test} dataset)?}

Table \ref{Methodology:RQ3B--removed-samples-with-multilple-CWEs} presents our analysis based on Table \ref{Methodology:VulRepair-dataset-Consistency}.

\begin{table}[h]
\centering
  \caption{RQ3B Dataset composition (removing cross-set duplicates and inconsistent samples from the test set)}
  \label{Methodology:RQ3B--removed-samples-with-multilple-CWEs}
  \begin{tabular}{|l|r|r|}
    \hline
    Samples &Train &Test\\
    \hline
    Unique Samples (US = SL - CS DUP)                    &4,918   &672\\
  \hline
\end{tabular}
\end{table}

\begin{table}[h]
\centering
  \caption{Model Performance on RQ3B}
  \label{Methodology:RQ3B-dataset-ModelPerformance}
  \begin{tabular}{|l|r|r|}
    \hline
    Models          & PP RQ3B & \% PP Replicated\\
    \hline
    VulRepair       & 10.27\% & 25.5\% (10.27/40.24)\\
    \hline
    CodeBERT        & 5.38\% & 16.2\% \enspace(5.38/33.20)\\
    \hline
    GraphCodeBERT   & 6.25\% & 16.2\% \enspace(6.25/38.51)\\
    \hline
    UniXcoder       & 6.18\% & 15.0\% \enspace(6.18/40.96)\\
  \hline
\end{tabular}
\end{table}

\textbf{Result:}
Table \ref{Methodology:RQ3B-dataset-ModelPerformance} summarizes the results of Table \ref{Appendices:RQ3B} and shows that just as was seen in the uniqueness experiments, providing more training data (4,918 vs. 3,995 samples) yielded performance increases in each of the models tested, but the performance is well below that yielded on the RQ2B dataset.

\begin{table*}
\centering
  \caption{Model Performance on RQ1 vs RQ2B vs RQ3B - Top10CWEs}
  \label{Experiment:RQ1-RQ3-RQ4dataset-ModelPerformance-Top10CWEs}
  \resizebox{\textwidth}{!}{
  \begin{tabular}{|r|c|r|r|r|r|r|r|r|r|r|r|r|r|r|r|r|}
    \hline
    Rank    & CWE      &\multicolumn{3}{c|}{Total Samples} &\multicolumn{3}{c|}{VulRepair} &\multicolumn{3}{c|}{CodeBERT}  &\multicolumn{3}{c|}{GraphCodeBERT}  &\multicolumn{3}{c|}{UniXcoder}\\
    \cline{3-17}
            &               &{\footnotesize RQ1}&{\footnotesize RQ2B}&{\footnotesize RQ3B}      &{\footnotesize RQ1}&{\footnotesize RQ2B}&{\footnotesize RQ3B}   &{\footnotesize RQ1}&{\footnotesize RQ2B}&{\footnotesize RQ3B}  &{\footnotesize RQ1}&{\footnotesize RQ2B}&{\footnotesize RQ3B}     &{\footnotesize RQ1}&{\footnotesize RQ2B}&{\footnotesize RQ3B}\\
    \hline
    1       & {\footnotesize CWE-787}       &53&33&22                &32.7\%&3.5\%&\cellcolor[gray]{0.8}2.3\%      &17.6\%&2.5\%&0.0\%     &24.8\%&2.5\%&0.0\%       &28.0\%&5.0\%&1.5\%\\
    \hline
    2       & {\footnotesize CWE-79}        &1&0&0                   &0\% &0\%&0\%            &0\% &0\%&0\%           &0\%&0\%&0\%              &0\%&0\%&0\%\\
    \hline
    3       & {\footnotesize CWE-125}       &170&68&67               &29.9\%&5.9\%&6.5\%      &30.8\%&10.8\%&8.7\%    &38.2\%&11.0\%&\cellcolor[gray]{0.8}10.5\%     &42.7\%&12.2\%&10.0\%\\
    \hline
    4       & {\footnotesize CWE-20}        &152&73&63               &44.8\%&13.2\%&\cellcolor[gray]{0.8}9.0\%    &33.9\%&7.3\%&6.1\%     &38.4\%&10.3\%&7.9\%      &39.0\%&9.6\%&7.7\%\\
    \hline
    5       & {\footnotesize CWE-78}        &3&1&1                   &22.2\%&0\%&0\%          &22.2\%&0\%&0\%         &33.3\%&0\%&0\%           &44.4\%&0\%&0\%\\
    \hline
    6       & {\footnotesize CWE-89}        &5&1&1                   &56.7\%&0\%&0\%          &40.0\%&0\%&0\%         &53.3\%&0\%&0\%           &56.7\%&0\%&0\%\\
    \hline
    7       & {\footnotesize CWE-416}       &55&29&17                &51.2\%&6.9\%&0\%        &40.9\%&5.8\%&0\%       &48.8\%&9.2\%&0\%         &48.8\%&6.3\%&0\%\\
    \hline
    8       & {\footnotesize CWE-22}        &8&2&2                   &33.3\%&0\%&0\%          &29.2\%&0\% &0\%        &33.3\%&0\%&0\%           &35.4\%&0\%&0\%\\
    \hline
    9       & {\footnotesize CWE-352}       &2&2&2                   &0\%&0\%&0\%             &0\% &0\%&0\%           &0\%&0\%&0\%              &0\%&0\%&0\%\\
    \hline
    10      & {\footnotesize CWE-434}       &-&-&-                   &-&-&-                   &-&-&-                  &-&- &-                   &-&- &-\\
    \hline
            &                               &449&209&175             &38.0\%&8.05\%&6.0\%     &31.3\%&7.3\% &5.5\%    &37.8\%&8.85\%&\cellcolor[gray]{0.8}6.9\%      &40.2\%&9.0\% &6.8\%\\   
  \hline
\end{tabular}
}
\end{table*}

\subsection{Revisiting the Top 10 CWEs}

One of the interesting and compelling results in the initial VulRepair paper\cite{fu2022vulrepair} is a table discussing performance of VulRepair on CWE samples appearing in the 2021 comparitech list of the top 10 CWEs\cite{top10}.
Fu et al. reported that 38\% of the 449 such samples in the VulRepair dataset were correctly repaired but we observed a substantial reduction in model performance, highlighting the critical role that these defective samples play in model effectiveness. We now turn to the question of how much the removal of inaccurate and incomplete samples will reduce the top 10 dataset.


\textbf{RQ4A: What effect does removing duplicate data from VulRepair have on the repair of vulnerabilities included in the top ten CWEs.}

We first revisit this question in Table \ref{Experiment:RQ1-RQ3-RQ4dataset-ModelPerformance-Top10CWEs}.
The detailed results are in \cref{Appendices:RQ1-RQ2B-CWE-787,Appendices:RQ1-RQ2B-CWE-125,Appendices:RQ1-RQ2B-CWE-20,Appendices:RQ1-RQ2B-CWE-78,Appendices:RQ1-RQ2B-CWE-89,Appendices:RQ1-RQ2B-CWE-416,Appendices:RQ1-RQ2B-CWE-22,Appendices:RQ1-RQ2B-CWE-352}.
We report the performance of each of the models studied on the top 10 CWEs, showing their performance when duplicate and inconsistent samples are removed from consideration.
First, we note the significant reduction in the number of top 10 CWE samples--only 39\% remain when the defective samples are removed.
But of even greater concern is the fact that across all four models, when these samples are removed the average performance drops to 
17\% of the original performance on average.
For instance, looking at the performance of CodeBERT on the RQ1 dataset, the model achieved 31.3\%. However, when evaluated on the RQ3B dataset, its performance fell to just 5.5\%, resulting in an average performance drop to 17\%. This highlights how heavily the samples introduced errors.

That means that the performance report on the top 10 CWEs was severely overestimated as a result of data leakage.

\textbf{RQ4B: How does identifying and analyzing inaccurate and incomplete samples of the top ten CWEs in the test set of the deduplicated VulRepair dataset change our understanding of its performance?}


Once duplicate samples are removed from the VulRepair dataset, many inaccurate and incomplete samples will remain.
We were able to address this question only in part, however, due to time constraints we were only able to consider 8 of the top 10 categories.

\subsubsection{Accuracy}\label{Result-Accuracy}
In this work, the primary determiner of accuracy is the correctness of the CWE label associated with a vulnerable sample.
Accuracy plays an essential role during model training and validation.
An incorrect label potentially leads the model to learn an incorrect pattern associated with given CWE type, thereby adversely affecting model evaluation.
The accuracy of sample labeling is largely determined by the methods used in data collection.
The VulRepair dataset collection process involved gathering data related to real-world vulnerabilities and tracing the sample to its vulnerability fix commit.
These links are derived from external vulnerability reports listed in databases such as the NVD.
This method assumes that each vulnerability fix commit is exclusively dedicated to addressing that vulnerability, overlooking potential confounding factors.
These factors include non-functional changes (such as stylistic modifications), tertiary alterations (such as adjustment of various variables and multiple functions in a way that is unrelated to the vulnerability), and tangled commits (where a single change addresses multiple issues simultaneously)\cite{herzig2013impact}.

In Table \ref{Experiment:RQ3Adataset-AccuracyPerformance-Top10CWEs} we show the number of samples from RQ2B that remain when only accurate samples are included in 8 of the top 10 CWEs.
Note that these inaccurate samples form a superset of inconsistent samples (such as are removed in RQ3B).
Removing inaccurate samples preserves only 44\% (30/68) of the unique samples.
Given the extremely small numbers in each category, the statistical importance of the results becomes much less apparent.

\begin{table*}
\centering
  \caption{RQ2B Data Quality - Top10CWEs}
  \label{Experiment:RQ3Adataset-AccuracyPerformance-Top10CWEs}
  \begin{tabular}{|l|l|l|r|r|r|r|}
    \hline
    Rank    & CWE Type & Name                          & RQ2B Samples   &Accurate &Complete & Accurate \& Complete\\
    \hline
    1       & CWE-787   & Out-of-bounds Write          &33              &15        &18      &12\\
    \hline
    2       & CWE-79    & Cross-site Scripting         &0               &0         &0       &0\\
    \hline
    5       & CWE-78    & OS Command Injection         &1               &0         &0       &0\\
    \hline
    6       & CWE-89    & SQL Injection                &1               & 1        &1       &1\\
    \hline
    7       & CWE-416   & Use After Free               &29              &11        &18      &7\\
    \hline
    8       & CWE-22    & Path Traversal               &2               & 1        &0       &0\\
    \hline
    9       & CWE-352   & Cross-Site Request Forgery   &2               & 2        &1       &1\\
    \hline
    10      & CWE-434   & Dangerous File Type          &-               &-         &-       &-\\
    \hline
            &           & Total                        &68              &30        &38      &21\\    
  \hline
\end{tabular}
\end{table*}



\subsubsection{Completeness}\label{Result-Completeness}
The concept of completeness is pivotal and pertains to the acquisition of comprehensive information necessary to identify and remedy vulnerabilities within a given dataset or individual CWE sample (VulRepair sample).
In some cases, not all of the code necessary to either understand or repair a vulnerability is available in the sample itself.
Such incomplete samples can adversely affect the learning efficacy of the model, thus leading to incomplete or ineffective model prediction.
During analysis, we observed that many samples lack relevant code related to the vulnerability repair.
These samples were subsequently classified as \emph{unverifiable samples}. Also discussed in Section \ref{Data-Attributes-Analysis}.

When incomplete samples are removed from the dataset, only 56\% (38/68) of samples remain as shown in Table \ref{Experiment:RQ3Adataset-AccuracyPerformance-Top10CWEs}.
Furthermore, when we remove those samples that are either inaccurate or incomplete, we are left with less that 1/3 of the unique samples, and only 16.5\% (21/127 for the 8 CWE categories) of the samples reported in Fu et al.\cite{fu2022vulrepair}
We would hesitate to try to draw significant conclusions about performance of a model from such a small collection of samples.

\subsection{Transfer Learning}\label{TransferLearning}
As discussed in the introduction and detailed in the related work, Fu et al.\cite{fu2024vision} did not utilize the bug-fix dataset provided by Chen et al.\cite{chen2022neural} 
The pre-training using the employed bug-fix dataset is compromised by a high number of in-set and cross-set duplicates, as well as significant overlap between the bug-fix pre-training dataset and the vulnerability fine-tuning dataset. 
This significantly undermines the purpose of pre-training on the bug-fix dataset before fine-tuning. This suggest that the results presented based on that training set require reevaluation.

Zhang et al.\cite{zhang2023pre} and Antonio et al.\cite{mastropaolo2024training} also present results of transfer learning in their respective papers, but both suffer from data leakage issues in pre-training and fine-tuning.
Both authors used the bug-fix dataset for pre-training, but it contains 6,192 duplicate entries in the training dataset, 4 duplicate entries in the validation dataset, and 247 cross-set duplicates.
Therefore that dataset needs to be cleaned before it can be effectively employed.
Furthermore, both authors used the VulRepair dataset, which is plagued by high duplication rates.
While Antonio et al. addressed cross-set duplicates they did not address in-set duplicates and inconsistently labeled samples leading to potential bias.
Thus, the  performance of transfer learning remains dubious.
All of these previously reported transfer learning results employed data sets with quality issues.

In this section, we evaluate the documented approach presented by Chen et al.\cite{chen2022neural}, which suggests that using a larger, generic bug-fix corpus can improve the performance of vulnerability repair models but only with datasets that are free from duplicate entries.
We have three training scenarios. The first scenario, termed the \emph{bug-fix scenario}, involves training the model solely on the de-duplicated bug-fix dataset of Chen et al.\cite{chen2022neural}. The hyperparameter settings for this scenario are described in \ref{Pre-training-hyperparameters}.
The second scenario, which we call the \emph{vulnerability repair scenario}, focuses exclusively on training using the RQ3B dataset.
The hyperparameter settings for this scenario are described in \ref{Fine-tuning-hyperparameters}.
The third scenario, the \emph{transfer learning scenario}, involves a two-step training process where the model is initially trained on the bug-fix dataset and subsequently fine-tuned on the RQ3B dataset. 
This structure allows us to evaluate model effectiveness under different training conditions and ascertain the benefits of transfer learning in enhancing model performance on vulnerability repair tasks.

In this section, we also analyze model performance across different beam sizes, which serves two critical purposes.
First, as highlighted in previous research, a smaller beam size is more feasible for manual inspection of the generated repairs\cite{fu2024vision,mastropaolo2024training}.
Second, this analysis provides a clearer picture of how the model performs at various beam sizes compared to prior studies, which helps pinpoint irregularities in reported model performance across different studies.
For instance, recent findings by VQM \cite{fu2024vision} report their new vision-transformer model achieves accuracy rates of 32.33\%, 42.72\%, and 45.14\% at beam sizes of 1, 3, and 5, respectively. In the same study, VulRepair attains accuracy rates of 29.65\%, 39.85\%, and 42.79\% at the corresponding beam sizes.
Because of the data quality issues in the bug-fix dataset model used to train them, these likely overestimate model performance.
In contrast, research by Antonio et al.\cite{mastropaolo2024training} reports lower performance for the VulRepair model, with accuracy scores of 12.28\%, 17.58\%, and 18.64\% at the same beam sizes. This discrepancy is almost surely due to data leakage in both the bug-fix and vulnerability repair datasets as well as inconsistently labeled samples in the vulnerability dataset.

\begin{table}[h]
\centering
  \caption{Bug-Fix and Transfer Learning Performance at Beam size of 1,3 and 5}
  \label{Experiment:Transfer-Learning-Beam1-5}
  \resizebox{\columnwidth}{!}{
  \begin{tabular}{|l|r|r|r|r|r|r|}
    \hline
    Models                  &\multicolumn{2}{c|}{Beam = 1}   &\multicolumn{2}{c|}{Beam = 3}   &\multicolumn{2}{c|}{Beam = 5} \\
    \cline{2-7}
                            &{BF}&{TL}                      &{BF}&{TL}                     &{BF}&{TL}\\
    \hline
    VulRepair               &3.6\%&13.5\%                  &7.4\%&19.0\%                &7.6\%&20.2\%\\
    \hline
    CodeBERT                &3.0\%&12.5\%                  &4.6\%&17.3\%                &5.4\%&18.9\%\\
    \hline
    GraphCodeBERT           &2.2\%&11.5\%                  &4.6\%&16.9\%                &5.8\%&19.0\%\\
    \hline
    UniXcoder               &1.9\%&12.9\%                  &5.2\%&18.1\%                &6.6\%&19.7\%\\
    \hline
    \multicolumn{7}{l}{Note: BF = Bug-Fix, TL = Transfer Learning}\\
\end{tabular}
}
\end{table}

Tables \ref{Experiment:Transfer-Learning-Beam1-5} present the results of bug-fix and transfer learning achieved by the models at beam sizes 1, 3, and 5. 
Detailed outcomes are reported in Table \ref{Appendices:RQ5-pretrain}. 
As anticipated, VulRepair exhibited a performance decline compared to the findings reported by Fu et al.\cite{fu2024vision} However, it achieved superior results relative to Antonio et al.\cite{mastropaolo2024training}.

Table \ref{Experiment:Transfer-Learning-Beam50} present the overall result of the models in three different scenarios: bug-fix training, vulnerability training, and transfer learning at beam size 50.


\begin{table}[h]
\centering
  \caption{Transfer Learning Performance at Beam size 50}
  \label{Experiment:Transfer-Learning-Beam50}
  \begin{tabular}{|l|r|r|r|}
    \hline
    Models                  &\multicolumn{3}{c|}{Beam = 50}\\
    \cline{2-4}
                            &{BF}&{VR}&{TL}\\
    \hline
    VulRepair               &6.55\%&10.27\%&18.67\%\\
    \hline
    CodeBERT                &11.76\%&5.38\%&24.55\%\\
    \hline
    GraphCodeBERT           &11.76\%&6.25\%&25.42\%\\
    \hline
    UniXcoder               &11.31\%&6.18\%&26.07\%\\
    \hline
    \multicolumn{4}{l}{Note: BF = Bug-Fix, VR = Vulnerability Repair}\\
    \multicolumn{4}{l}{and TL = Transfer Learning}\\
\end{tabular}
\end{table}

These results clearly show that pre-training with a large corpus can provide improved performance on security vulnerability repair, achieving results nearly as good as models trained on the de-duplicated VulRepair dataset.
But the more important result is that employing models pre-trained with bug-fix and using deduplicated VulRepair data for transfer learning lets each model achieve performance that is slightly better than that or our RQ3B experiment with a beam size of 50 even when a beam size of just 1 is used and more than twice that of RQ3B with a beam size of 5. As shown in Table \ref{Experiment:Transfer-Learning-Beam1-5}.

\begin{figure}[h]
   \centering
   \includegraphics[width=\linewidth]{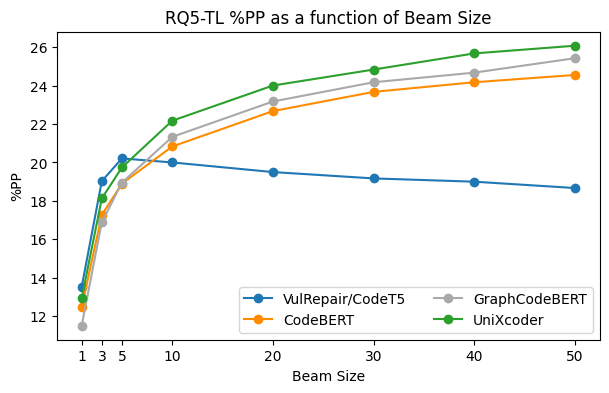}
   \caption{\%PP as a Function of Beam Size for RQ5 (Transfer Learning)}
   \label{Experiment:Transfer-Learning-Beam-1-50}
   \centering
 \end{figure}

To more fully understand the impact of beam size on performance of training the models with transfer learning, we evaluated these models using beam sizes in the range 1 to 50 as reported in \cref{Appendices:RQ5-finetrain-VulRepair,Appendices:RQ5-finetrain-CodeBERT,Appendices:RQ5-finetrain-GraphCodeBERT,Appendices:RQ5-finetrain-UniXcoder}.
Figure \ref{Experiment:Transfer-Learning-Beam-1-50} shows the results of this analysis.
It demonstrates that VulRepair/CodeT5 reaches its best performance around beam size 5, whereas the rest of the models show diminishing performance improvements thereafter.
This behavior is different from that reported in Fu et al.\cite{fu2022vulrepair} and our experiments RQ1 through RQ3B where performance increases through beam size 50.
We are uncertain of the cause of this change in behavior.
Choosing an appropriate beam size for program repair represents a trade-off between programmer effort and possible better performance.
This work suggests that a beam size somewhere between 5 and 10 may be most appropriate for the task of program repair.


\section{Threats to Validity}
As the goal of this paper is to show how the effect of dataset quality issues may have caused inappropriate estimates of model performance and this work is aimed at comparing reported performance to that of the same models with dataset quality issues removed, the primary threats to the validity of our reported results arise from two issues:

\begin{enumerate}
    \item The model implementations we have employed may be faulty.
    \item The dataset quality improvements we have applied may be flawed.
\end{enumerate}

We have taken precautions against model implementation errors by employing the referenced authors' websites for  VulRepair and CodeBERT\cite{vulrepairgit} as well as  GraphCodeBERT and UniXCoder\cite{isenglabgit}.
Minimal changes were made to the code, namely
\begin{enumerate}
    \item The CodeBERT model was modified to generate output to a CSV file.
    \item The number of GPUs for GraphCodeBERT and UniXcoder were modified for RQ1, RQ2A, RQ2B, RQ3A, RQ3B, and RQ5.
\end{enumerate}

The datasets we have used by the previous authors are available from the sources cited in their respective works.
Determination of label accuracy and completeness for results on the top 10 CWEs was initially carried out by author 1 and those samples that raised questions were also inspected by author 2.
We are open to the possibility that errors may have occurred in this analysis.
That is why all the data and analysis notes are available from our github site.

The more compelling issue associated with data quality in the VulRepair dataset, concerns the issues of accuracy and completeness.
The work reported in Sections \ref{Result-Accuracy} and \ref{Result-Completeness} regarding accuracy and completeness only partially identifies and analyzes inaccurate and incomplete samples from the RQ2B test set samples that were used to create results for the referenced experiments.
The amount of labor required to scour the entire dataset for such defective samples could not be carried out in time for this presentation.
It is certain that numerous inaccurate and incomplete samples persist in that dataset's training, validation, and test samples.
The full extent of the effect of these samples on model performance is still unknown.


\section{Related Work} \label{Related-Work}
Mastropaolo et al.\cite{mastropaolo2024training} investigate different training strategies for vulnerability repair.
They begin by replicating the VulRepair work, identifying issues with cross-set duplicates in the VulRepair dataset.
However, they were unable to fully resolve the problem due to the presence of unresolved duplicates with label mismatches, i.e., inconsistent labels (duplicate code with conflicting CWE tags).
This resulted in 259 cross-set duplicate samples in the test set and 49 in the validation set overlapping with the training set. 
Additionally, the dataset contains in-file duplicates, comprising 1,424 in the training set, 35 in the validation set, and 109 in the test set. 
In total, the dataset suffers from 1,424 duplicates in the training set, 84 in the validation set, and 368 in the test set.


Test set duplicates can skew performance estimates either positively or negatively, depending on whether the duplicated instances are correctly or incorrectly repaired. 
Validation set duplicates may lead to premature termination of the training process or cause it to run longer than necessary, contingent on the correctness of the repairs for these duplicates.
Training set duplicates introduce a slight bias, preferentially enhancing the model's performance on duplicated entries compared to non-duplicated ones.
These issues are addressed in Sections \ref{Uniqueness} and \ref{Consistency} of our paper.
While Mastropaolo et al. limit their evaluation to CodeT5, we extend the analysis by including CodeBERT, GraphCodeBERT, and UniXCoder, demonstrating their superior performance compared to CodeT5 at beam sizes of 10 or more.
Furthermore, unlike Mastropaolo et al.\cite{mastropaolo2024training}, we performed a manual analysis of the top 10 CWE categories in the dataset, verifying the accuracy and completeness of the samples, which is crucial for understanding the characteristics of vulnerability samples.
Our findings also indicate that transfer learning outperforms the prompt tuning method used by the authors.

Ding et al.\cite{ding2024vulnerability} similarly highlight issues with exact code duplicates and low label accuracy in datasets like CVEfixes and BigVul.
Their focus, however, is on the effectiveness of large language models (LLMs) for vulnerability detection, which is distinct from program repair.
Their work lacks a comprehensive evaluation of models on deduplicated datasets, instead proposing a new method for constructing a robust dataset and evaluating models on it.
Their proposed dataset is not well-suited for program repair as many samples lack CWE types which are essential for characterizing vulnerability samples and training repair models.

Fu et al.\cite{fu2024vision} reported further work on program repair by introducing a vision transformer model leveraging the bug-fix provided by VRepair\cite{chen2022neural} and vulnerability datasets (VQM).
The authors claim that VRepair was initially pre-trained on a bug-fix dataset of 23,607 C/C++ functions.
In fact, as noted in Section \ref{Introduction}, VRepair was pre-trained on a bug-fix dataset with 534,848 training and 10,000 validation samples\cite{chen2022neural}.
Our analysis of the bug-fix dataset\cite{vqm2024} used for VQM pre-training revealed that it contains 21,246 training samples and 2,362 validation samples.
We identified 17,303 duplicate entries in the training set and 666 duplicates in the validation set, reducing the dataset to 3,943 unique training samples and 1,695 unique validation samples.
Following the removal of in-set duplicates, we also performed a cross-set duplicate analysis, which identified 1,613 duplicates shared between the training and validation sets.
During manual verification, we noticed several samples appeared identical except for misplaced \texttt{<S2SV\_StartBug>} and \texttt{<S2SV\_EndBug>} tags, which the Pandas \texttt{drop\_duplicates()} will not identify as matches even though their code is identical.
After removing these tags and reanalyzing the data, we found an extra 1319 (18,622 - 17303) duplicate entries in the training set and 116 (782 - 666) duplicates in the validation set, leaving 2,624 unique training samples and 1,579 unique validation samples.
A subsequent cross-set duplicate check revealed that \emph{all} of the validation set code samples are present in the training set, suggesting significant data overlap.
Additionally, we observed that the pre-tag of the samples in the bug-fix dataset includes vulnerability CWE tags, indicating that these samples belong to vulnerability categories rather than generic bug-fix categories (CWE-000).
This observation led us to suspect potential data leakage between the bug-fix and fine-tuned vulnerability datasets (VQM).
To investigate further, we conducted a cross-set duplicate analysis by comparing the test, validation, and training sets from the fine-tuned vulnerability dataset against the training set of the bug-fix dataset used by VQM.
Our analysis revealed 511 matching entries out of 1,090 test set samples, 243 matches from the 536 validation set samples, and 1,747 duplicates from the 3,790 training set samples in the bug-fix dataset's training set.
These results demonstrate a significant overlap between the datasets.
These findings raise concerns about data leakage, which could have contributed to the reported performance results of 32.33\%, 42.72\%, and 45.14\% at beam sizes of 1, 3, and 5, respectively.
Given this overlap, we contend that retraining the model without these duplicates is necessary to assess its true performance accurately.

Another recent work, DiverseVul\cite{chen2023diversevul}, is a vulnerability dataset that follows an automated labeling approach, collecting vulnerability-fixing commits from open-source databases such as NVD, assuming that all such commits modify only the vulnerable function.
This approach introduces significant label noise, however, with the dataset being reported as only 60\% accurate.
Furthermore, this dataset does not include repaired code and is unsuitable for use in vulnerability repair systems.
\section{Conclusion}
\label{Conclusion}

When one trains a model to correctly repair program vulnerabilities, its accuracy depends on several critical factors concerning data employed in training or fine-tuning.
First, the presence of the actual vulnerability within each buggy program sample is essential.
Second, each buggy sample program must be accurately described using a Common Weakness Enumeration (CWE) label that correctly identifies the type of vulnerability.
Additionally, for each sample needing repair, it is crucial that the \texttt{<StartBug>} and \texttt{<EndBug>} tags are precisely placed to assist the model in recognizing and understanding the vulnerability.
Finally, when validating the model's output, it is imperative to compare it against a reference code that is complete and contains no missing tokens.
These components are vital to ensuring that the model's predictions are both accurate and reliable.
We observed that a significant number of samples in various datasets lack one or more of these criteria, adversely affecting the overall performance of vulnerability repair models.

The work we reported was verified over multiple runs of models trained with different random seeds.
When reporting model performance, it is crucial to report the mean accuracy because significant variations in performance are observed even when testing on the same dataset. 

Our work has replicated the results of others using datasets with uniqueness and consistency defects removed.
The results we achieved removing only these two kinds of problems make is clear that previous studies have overestimated the capabilities of the models.
We demonstrated that simply pre-training a model with a large corpus of bug-fix data can yield performance similar to the actual performance of the reported systems on a deduplicated VulRepair dataset and that employing transfer learning can improve them dramatically more.
We have seen nothing that proves that automated program repair is beyond the capability of LLMs, only that the datasets we have used to train and test such systems are hopelessly small and have been riddled with errors.

We showed that inconsistently labeled samples, repeating the same buggy and repaired code with different CWE tags, yield over-estimated performance similar to duplicate samples.
All of the models tested showed the impact of data leakage from such samples.

The value of high-quality, large datasets to support program repair cannot be overstated.
Collecting and curating these datasets is a monumental task.
The ACM artifact review and badging effort is a good start, but if we are to address the needs of the security vulnerability repair community, it needs to be expanded.
The fact that the primary dataset studied here received an \emph{Artifacts Evaluated-Reusable} badge (as discussed in Section \ref{Dataset:Discussion}) indicates both the difficulty of such evaluations and the need to take extreme care.
Such datasets should be the property and responsibility of the community with well-defined methods for contributions and public auditing.
Only when we can trust our data can we trust our results.



\appendices
\onecolumn
\section{DETAILED RESULTS}
\begin{table}[H]
\centering
  \caption{Performance Of Models on RQ1}
  \label{Appendices:RQ1}
  \begin{tabular}{|l|r|r|r|r|r|r|r|}
    \hline
    Models          & Seed 26312 & Seed 43511 & Seed 67732 & Seed 70757 & Seed 95541 & Seed 123456 & Average PP\\
    \hline
    VulRepair       & 0.4349    & 0.4197 & 0.4244 & 0.3511 & 0.4244 & 0.371  & 40.42\% (2.4255/6)\\
    \hline
    CodeBERT        & 0.3294    & 0.3687 & 0.3406 & 0.2327 & 0.3992 & 0.3212 & 33.20\% (1.9918/6)\\
    \hline
    GraphCodeBERT   & 0.4238    & 0.3904 & 0.3576 & 0.3857 & 0.3804 & 0.3728 & 38.51\% (2.3107/6)\\
    \hline
    UniXcoder       & 0.4045    & 0.4174 & 0.4467 & 0.3986 & 0.3499 & 0.4402 & 40.96\% (2.4573/6)\\
    \hline
  \end{tabular}
\end{table}
\begin{table}[H]
\centering
  \caption{Performance Of Models on RQ2A}
  \label{Appendices:RQ2A}
  \begin{tabular}{|l|r|r|r|r|r|r|r|}
    \hline
    Models          & Seed 26312 & Seed 43511 & Seed 67732 & Seed 70757 & Seed 95541 & Seed 123456 & Average PP\\
    \hline
    VulRepair       & 0.096  & 0.1053 & 0.0786 & 0.1034 & 0.0848 & 0.0663 & 8.91\% (0.5344/6)\\
    \hline
    CodeBERT        & 0.0545 & 0.0514 & 0.0588 & 0.0471 & 0.0588 & 0.0644 & 5.58\% (0.3350/6)\\
    \hline
    GraphCodeBERT   & 0.057  & 0.0495 & 0.0545 & 0.0508 & 0.0514 & 0.0551 & 5.31\% (0.3183/6)\\
    \hline
    UniXcoder       & 0.0601 & 0.0557 & 0.0483 & 0.0477 & 0.0378 & 0.0396 & 4.82\% (0.2892/6)\\
    \hline
  \end{tabular}
\end{table}
\begin{table}[H]
\centering
  \caption{Performance Of Models on RQ2B}
  \label{Appendices:RQ2B}
  \begin{tabular}{|l|r|r|r|r|r|r|r|}
    \hline
    Models          & Seed 26312 & Seed 43511 & Seed 67732 & Seed 70757 & Seed 95541 & Seed 123456 & Average PP\\
    \hline
    VulRepair       & 0.1551  & 0.0965 & 0.1575 & 0.1526 & 0.1111 & 0.1172 & 13.17\% (0.79/6)\\
    \hline
    CodeBERT        & 0.1038  & 0.0769 & 0.0916 & 0.0891 & 0.0842 & 0.0842 & 8.83\% (0.5298/6)\\
    \hline
    GraphCodeBERT   & 0.1001  & 0.0965 & 0.094  & 0.0916 & 0.083  & 0.0879 & 9.22\% (0.5531/6)\\
    \hline
    UniXcoder       & 0.083   & 0.1197 & 0.1062 & 0.0794 & 0.0904 & 0.0672 & 9.10\% (0.5459/6)\\
    \hline
  \end{tabular}
\end{table}
\begin{table}[H]
\centering
  \caption{Performance Of Models on RQ3A}
  \label{Appendices:RQ3A}
  \begin{tabular}{|l|r|r|r|r|r|r|r|}
    \hline
    Models          & Seed 26312 & Seed 43511 & Seed 67732 & Seed 70757 & Seed 95541 & Seed 123456 & Average PP\\
    \hline
    VulRepair       & 0.0834  & 0.074  & 0.0777  & 0.0445  & 0.0665  & 0.0821   & 7.14\% (0.4282/6)\\
    \hline
    CodeBERT        & 0.0426  & 0.042  & 0.0326  & 0.0307  & 0.032   & 0.0357   & 3.59\% (0.2156/6)\\
    \hline
    GraphCodeBERT   & 0.0376  & 0.0364 & 0.0351  & 0.042   & 0.0376  & 0.0364   & 3.75\% (0.2251/6)\\
    \hline
    UniXcoder       & 0.0426  & 0.042  & 0.0364  & 0.0445  & 0.037   & 0.0439   & 4.11\% (0.2464/6)\\
    \hline
  \end{tabular}
\end{table}
\begin{table}[H]
\centering
  \caption{Performance Of Models on RQ3B}
  \label{Appendices:RQ3B}
  \begin{tabular}{|l|r|r|r|r|r|r|r|}
    \hline
    Models          & Seed 26312 & Seed 43511 & Seed 67732 & Seed 70757 & Seed 95541 & Seed 123456 & Average PP\\
    \hline
    VulRepair       & 0.1012  & 0.0878 & 0.0923  & 0.1071 & 0.1101 & 0.1176 & 10.27\% (0.6161/6)\\
    \hline
    CodeBERT        & 0.0521  & 0.0506 & 0.058   & 0.0551 & 0.0565 & 0.0506 & 5.38\% (0.3229/6)\\
    \hline
    GraphCodeBERT   & 0.0625  & 0.0506 & 0.0685  & 0.058  & 0.067  & 0.0685 & 6.25\% (0.3751/6)\\
    \hline
    UniXcoder       & 0.061   & 0.0625 & 0.0685  & 0.0685 & 0.0565 & 0.0536 & 6.18\% (0.3706/6)\\
    \hline
  \end{tabular}
\end{table}
\makeatletter
\newcommand\jnwtiny{\@setfontsize\jnwtiny{6}{7}}
\makeatother
\begin{table}[H]
\centering
  \caption{Model Performance on RQ1, RQ2B and RQ3B - CWE-787}
  \label{Appendices:RQ1-RQ2B-CWE-787}
  \resizebox{\columnwidth}{!}{
  \begin{tabular}{|l|r|r|r|r|r|r|r|r|r|r|r|r|r|r|r|r|r|r|r|r|r|}
    \hline
    Model & \multicolumn{3}{c|}{Seed 26312} & \multicolumn{3}{c|}{Seed 43511} & \multicolumn{3}{c|}{Seed 67732} & \multicolumn{3}{c|}{Seed 70757} & \multicolumn{3}{c|}{Seed 95541} & \multicolumn{3}{c|}{Seed 123456} & \multicolumn{3}{c|}{Average PP}\\
    \cline{2-22}
    & {\tiny RQ1} & {\tiny RQ2B} & {\tiny RQ3B}
    & {\tiny RQ1} & {\tiny RQ2B} & {\tiny RQ3B}
    & {\tiny RQ1} & {\tiny RQ2B} & {\tiny RQ3B}
    & {\tiny RQ1} & {\tiny RQ2B} & {\tiny RQ3B}
    & {\tiny RQ1} & {\tiny RQ2B} & {\tiny RQ3B}
    & {\tiny RQ1} & {\tiny RQ2B} & {\tiny RQ3B} 
    & {\tiny RQ1} & {\tiny RQ2B} & {\tiny RQ3B}\\
    \hline
    VR  
        &22 & 1  & 1
        &18 & 0  & 0
        &18 & 1  & 0
        &12 & 2  & 0
        &17 & 2  & 1
        &17 & 1  & 1
        &32.7\%& 3.5\%  & 2.3\%\\
        \hline
    CB 
        &11 & 1 &  0
        &8  & 0 &  0
        &8  & 2 &  0
        &6  & 0 &  0
        &14 & 1 &  0
        &9  & 1 &  0
        &17.6\% & 2.5\%  & 0.0\%\\
        \hline
    GCB 
        &16 & 1 & 0  
        &14 & 1 & 0
        &12 & 1 & 0
        &14 & 1 & 0 
        &11 & 1 & 0
        &12 & 0 & 0
        &24.8\% & 2.5\% & 0.0\%\\
        \hline
    UX     
        &15 & 2 & 0  
        &16 & 2 & 1 
        &18 & 2 & 1
        &12 & 1 & 0
        &10 & 2 & 0
        &18 & 1 & 0
        &28.0\%  & 5.0\%  & 1.5\%\\
    \hline
    \multicolumn{20}{l}{Note: RQ1, RQ2B, and RQ3B present total correct predictions out of 53, 33, and 22 total samples, respectively.}\\
  \end{tabular}
  }
\end{table}
\begin{table}[H]
\centering
  \caption{Model Performance on RQ1, RQ2B and RQ3B - CWE-125}
  \label{Appendices:RQ1-RQ2B-CWE-125}
  \resizebox{\columnwidth}{!}{
  \begin{tabular}{|l|r|r|r|r|r|r|r|r|r|r|r|r|r|r|r|r|r|r|r|r|r|}
    \hline
    Model & \multicolumn{3}{c|}{Seed 26312} & \multicolumn{3}{c|}{Seed 43511} & \multicolumn{3}{c|}{Seed 67732} & \multicolumn{3}{c|}{Seed 70757} & \multicolumn{3}{c|}{Seed 95541} & \multicolumn{3}{c|}{Seed 123456} & \multicolumn{3}{c|}{Average PP}\\
    \cline{2-22}
    & {\tiny RQ1} & {\tiny RQ2B} & {\tiny RQ3B}
    & {\tiny RQ1} & {\tiny RQ2B} & {\tiny RQ3B}
    & {\tiny RQ1} & {\tiny RQ2B} & {\tiny RQ3B}
    & {\tiny RQ1} & {\tiny RQ2B} & {\tiny RQ3B}
    & {\tiny RQ1} & {\tiny RQ2B} & {\tiny RQ3B}
    & {\tiny RQ1} & {\tiny RQ2B} & {\tiny RQ3B} 
    & {\tiny RQ1} & {\tiny RQ2B} & {\tiny RQ3B}\\
    \hline
    VR   
        &52 & 5 & 4
        &54 & 3 & 3
        &57 & 4 & 4 
        &43 & 6 & 5
        &54 & 3 & 5
        &45 & 3 & 5
        &29.9\% & 5.9\% & 6.5\%\\
        \hline
    CB   
        &49 & 8 & 6
        &58 & 8 & 6
        &61 & 6 & 5
        &36 & 7 & 6
        &60 & 8 & 6
        &50 & 7 & 6
        &30.8\% & 10.8\% & 8.7\%\\
        \hline
    GDB 
        &71 & 7 & 6
        &66 & 7 & 6
        &56 & 9 & 8
        &67 & 7 & 7
        &65 & 7 & 7
        &65 & 8 & 8
        &38.2\% & 11.0\% & 10.5\%\\
        \hline
    UX   
        &75 & 9 & 7
        &74 & 8 & 6
        &79 & 11& 7
        &73 & 7 & 7
        &57 & 8 & 7
        &77 & 7 & 6
        &42.7\%  & 12.2\%  & 9.95\%\\
    \hline
    \multicolumn{21}{l}{Note: RQ1, RQ2B, and RQ3B present total correct predictions out of 170, 68, and 67 total samples, respectively.}\\
  \end{tabular}
  }
\end{table}
\begin{table}[H]
\centering
  \caption{Model Performance on RQ1, RQ2B and RQ3B - CWE-20}
  \label{Appendices:RQ1-RQ2B-CWE-20}
  \resizebox{\columnwidth}{!}{
  \begin{tabular}{|l|r|r|r|r|r|r|r|r|r|r|r|r|r|r|r|r|r|r|r|r|r|}
    \hline
    Model & \multicolumn{3}{c|}{Seed 26312} & \multicolumn{3}{c|}{Seed 43511} & \multicolumn{3}{c|}{Seed 67732} & \multicolumn{3}{c|}{Seed 70757} & \multicolumn{3}{c|}{Seed 95541} & \multicolumn{3}{c|}{Seed 123456} & \multicolumn{3}{c|}{Average PP}\\
    \cline{2-22}
    & {\tiny RQ1} & {\tiny RQ2B} & {\tiny RQ3B}
    & {\tiny RQ1} & {\tiny RQ2B} & {\tiny RQ3B}
    & {\tiny RQ1} & {\tiny RQ2B} & {\tiny RQ3B}
    & {\tiny RQ1} & {\tiny RQ2B} & {\tiny RQ3B}
    & {\tiny RQ1} & {\tiny RQ2B} & {\tiny RQ3B}
    & {\tiny RQ1} & {\tiny RQ2B} & {\tiny RQ3B} 
    & {\tiny RQ1} & {\tiny RQ2B} & {\tiny RQ3B}\\
    \hline
    VR  
        &73 & 12 & 5
        &68 & 8  & 4
        &73 & 11 & 5
        &61 & 9  & 9
        &71 & 10 & 6
        &63 & 8  & 5
        &44.8\% & 13.2\% & 8.99\%\\
        \hline
    CB        
        &52 & 7 & 5
        &58 & 5 & 3
        &50 & 6 & 5
        &39 & 5 & 2
        &61 & 5 & 4
        &49 & 4 & 4
        &33.9\%  & 7.3\%  & 6.1\%\\
        \hline
    GDB 
        &63 & 8 & 5
        &58 & 8 & 4
        &56 & 7 & 5
        &58 & 6 & 5
        &57 & 9 & 6
        &58 & 7 & 5
        &38.4\% & 10.3\%  & 7.9\%\\
        \hline
    UX     
        &56 & 7 & 6
        &60 & 10 & 6
        &64 & 8  & 6
        &57 & 6  & 4
        &54 & 7  & 4
        &65 & 4  & 3
        &39.0\% & 9.6\%  & 7.7\%\\
    \hline
    \multicolumn{20}{l}{Note: RQ1, RQ2B, and RQ3B present total correct predictions out of 152, 73, and 63 total samples, respectively.}\\
  \end{tabular}
  }
\end{table}
\begin{table}[H]
\centering
  \caption{Model Performance on RQ1, RQ2B and RQ3B - CWE-78}
  \label{Appendices:RQ1-RQ2B-CWE-78}
  \resizebox{\columnwidth}{!}{
  \begin{tabular}{|l|r|r|r|r|r|r|r|r|r|r|r|r|r|r|r|r|r|r|r|r|r|}
    \hline
    Model & \multicolumn{3}{c|}{Seed 26312} & \multicolumn{3}{c|}{Seed 43511} & \multicolumn{3}{c|}{Seed 67732} & \multicolumn{3}{c|}{Seed 70757} & \multicolumn{3}{c|}{Seed 95541} & \multicolumn{3}{c|}{Seed 123456} & \multicolumn{3}{c|}{Average PP}\\
    \cline{2-22}
    & {\tiny RQ1} & {\tiny RQ2B} & {\tiny RQ3B}
    & {\tiny RQ1} & {\tiny RQ2B} & {\tiny RQ3B}
    & {\tiny RQ1} & {\tiny RQ2B} & {\tiny RQ3B}
    & {\tiny RQ1} & {\tiny RQ2B} & {\tiny RQ3B}
    & {\tiny RQ1} & {\tiny RQ2B} & {\tiny RQ3B}
    & {\tiny RQ1} & {\tiny RQ2B} & {\tiny RQ3B} 
    & {\tiny RQ1} & {\tiny RQ2B} & {\tiny RQ3B}\\
    \hline
    VR  
        &1 & 0 & 0
        &0 & 0 & 0
        &1 & 0 & 0
        &0 & 0 & 0
        &1 & 0 & 0
        &1 & 0 & 0
        &22.2\%  & 0\% & 0\%\\
        \hline
    CB   
        &1 & 0 & 0
        &1 & 0 & 0
        &0 & 0 & 0
        &0 & 0 & 0
        &1 & 0 & 0
        &1 & 0 & 0
        &22.2\% & 0\%  & 0\%\\
        \hline
    GDB  
        &1 & 0 & 0
        &2 & 0 & 0
        &1 & 0 & 0
        &1 & 0 & 0
        &1 & 0 & 0
        &0 & 0 & 0
        &33.3\% & 0\% & 0\%\\
        \hline
    UX  
        &2 & 0 & 0
        &2 & 0 & 0
        &1 & 0 & 0
        &1 & 0 & 0
        &1 & 0 & 0
        &1 & 0 & 0
        &44.4\%  & 0\%  & 0\%\\
    \hline
    \multicolumn{20}{l}{Note: RQ1, RQ2B, and RQ3B present total correct predictions out of 3, 1, and 1 total samples, respectively.}\\
  \end{tabular}
  }
\end{table}
\begin{table}[H]
\centering
  \caption{Model Performance on RQ1, RQ2B and RQ3B - CWE-89}
  \label{Appendices:RQ1-RQ2B-CWE-89}
  \resizebox{\columnwidth}{!}{
  \begin{tabular}{|l|r|r|r|r|r|r|r|r|r|r|r|r|r|r|r|r|r|r|r|r|r|}
    \hline
    Model & \multicolumn{3}{c|}{Seed 26312} & \multicolumn{3}{c|}{Seed 43511} & \multicolumn{3}{c|}{Seed 67732} & \multicolumn{3}{c|}{Seed 70757} & \multicolumn{3}{c|}{Seed 95541} & \multicolumn{3}{c|}{Seed 123456} & \multicolumn{3}{c|}{Average PP}\\
    \cline{2-22}
    & {\tiny RQ1} & {\tiny RQ2B} & {\tiny RQ3B}
    & {\tiny RQ1} & {\tiny RQ2B} & {\tiny RQ3B}
    & {\tiny RQ1} & {\tiny RQ2B} & {\tiny RQ3B}
    & {\tiny RQ1} & {\tiny RQ2B} & {\tiny RQ3B}
    & {\tiny RQ1} & {\tiny RQ2B} & {\tiny RQ3B}
    & {\tiny RQ1} & {\tiny RQ2B} & {\tiny RQ3B} 
    & {\tiny RQ1} & {\tiny RQ2B} & {\tiny RQ3B}\\
    \hline
    VR  
        &2 & 0 & 0
        &3 & 0 & 0
        &3 & 0 & 0
        &3 & 0 & 0
        &3 & 0 & 0
        &3 & 0 & 0
        &56.7\% & 0\%  & 0\%\\
        \hline
    CB 
        &2 & 0 & 0
        &1 & 0 & 0
        &2 & 0 & 0
        &2 & 0 & 0
        &3 & 0 & 0
        &2 & 0 & 0
        &40.0\% & 0\%  & 0\%\\
        \hline
    GDB  
        &3 & 0 & 0
        &3 & 0 & 0
        &2 & 0 & 0
        &3 & 0 & 0
        &3 & 0 & 0
        &2 & 0 & 0
        &53.3\%  & 0\%  & 0\%\\
        \hline
    UX 
        &3 & 0 & 0
        &3 & 0 & 0
        &3 & 0 & 0
        &3 & 0 & 0
        &2 & 0 & 0
        &3 & 0 & 0
        &56.7\% & 0\% & 0\%\\
    \hline
    \multicolumn{20}{l}{Note: RQ1, RQ2B, and RQ3B present total correct predictions out of 5, 1, and 1 total samples, respectively.}\\
  \end{tabular}
  }
\end{table}
\begin{table}[H]
\centering
  \caption{Model Performance on RQ1, RQ2B and RQ3B - CWE-416}
  \label{Appendices:RQ1-RQ2B-CWE-416}
  \resizebox{\columnwidth}{!}{
  \begin{tabular}{|l|r|r|r|r|r|r|r|r|r|r|r|r|r|r|r|r|r|r|r|r|r|}
    \hline
    Model & \multicolumn{3}{c|}{Seed 26312} & \multicolumn{3}{c|}{Seed 43511} & \multicolumn{3}{c|}{Seed 67732} & \multicolumn{3}{c|}{Seed 70757} & \multicolumn{3}{c|}{Seed 95541} & \multicolumn{3}{c|}{Seed 123456} & \multicolumn{3}{c|}{Average PP}\\
    \cline{2-22}
    & {\tiny RQ1} & {\tiny RQ2B} & {\tiny RQ3B}
    & {\tiny RQ1} & {\tiny RQ2B} & {\tiny RQ3B}
    & {\tiny RQ1} & {\tiny RQ2B} & {\tiny RQ3B}
    & {\tiny RQ1} & {\tiny RQ2B} & {\tiny RQ3B}
    & {\tiny RQ1} & {\tiny RQ2B} & {\tiny RQ3B}
    & {\tiny RQ1} & {\tiny RQ2B} & {\tiny RQ3B} 
    & {\tiny RQ1} & {\tiny RQ2B} & {\tiny RQ3B}\\
    \hline
    VR   
        &33 & 4 & 0
        &31 & 1 & 0
        &30 & 2 & 0
        &23 & 2 & 0
        &28 & 2 & 0
        &24 & 1 & 0
        &51.2\% & 6.9\%  & 0\%\\
        \hline
    CB     
        &21 & 3 & 0
        &24 & 0 & 0
        &23 & 3 & 0
        &11 & 2 & 0
        &30 & 1 & 0
        &26 & 1 & 0
        &40.9\%  & 5.8\% & 0\%\\
        \hline
    GDB
        &30 & 5 & 0
        &27 & 4 & 0
        &25 & 3 & 0
        &27 & 2 & 0
        &27 & 0 & 0
        &25 & 2 & 0
        &48.8\% & 9.2\% & 0\%\\
        \hline
    UX 
        &26 & 1 & 0
        &27 & 5 & 0
        &29 & 4 & 0
        &27 & 0 & 0
        &23 & 1 & 0
        &29 & 0 &  0
        &48.8\% & 6.3\%  & 0\%\\
    \hline
    \multicolumn{20}{l}{Note: RQ1, RQ2B, and RQ3B present total correct predictions out of 55, 29, and 17 total samples, respectively.}\\
  \end{tabular}
  }
\end{table}
\begin{table}[H]
\centering
  \caption{Model Performance on RQ1, RQ2B and RQ3B - CWE-22}
  \label{Appendices:RQ1-RQ2B-CWE-22}
  \resizebox{\columnwidth}{!}{
  \begin{tabular}{|l|r|r|r|r|r|r|r|r|r|r|r|r|r|r|r|r|r|r|r|r|r|}
    \hline
    Model & \multicolumn{3}{c|}{Seed 26312} & \multicolumn{3}{c|}{Seed 43511} & \multicolumn{3}{c|}{Seed 67732} & \multicolumn{3}{c|}{Seed 70757} & \multicolumn{3}{c|}{Seed 95541} & \multicolumn{3}{c|}{Seed 123456} & \multicolumn{3}{c|}{Average PP}\\
    \cline{2-22}
    & {\tiny RQ1} & {\tiny RQ2B} & {\tiny RQ3B}
    & {\tiny RQ1} & {\tiny RQ2B} & {\tiny RQ3B}
    & {\tiny RQ1} & {\tiny RQ2B} & {\tiny RQ3B}
    & {\tiny RQ1} & {\tiny RQ2B} & {\tiny RQ3B}
    & {\tiny RQ1} & {\tiny RQ2B} & {\tiny RQ3B}
    & {\tiny RQ1} & {\tiny RQ2B} & {\tiny RQ3B} 
    & {\tiny RQ1} & {\tiny RQ2B} & {\tiny RQ3B}\\
    \hline
    VR   
        &3 & 0 & 0
        &3 & 0 & 0
        &3 & 0 & 0
        &2 & 0 & 0
        &3 & 0 & 0
        &2 & 0 & 0
        &33.3\% & 0\%  & 0\%\\
        \hline
    CB  
        &2 & 0 & 0
        &3 & 0 & 0
        &2 & 0 & 0
        &1 & 0 & 0
        &3 & 0 & 0
        &3 & 0 & 0
        &29.2\% & 0\%  & 0\%\\
        \hline
    GDB
       &3 & 0 & 0
       &3 & 0 & 0
       &2 & 0 & 0
       &3 & 0 & 0
       &3 & 0 & 0
       &2 & 0 & 0
       &33.3\%  & 0\%  & 0\%\\
       \hline
    UX 
        &3 & 0 & 0
        &3 & 0 & 0
        &3 & 0 & 0
        &3 & 0 & 0
        &2 & 0 & 0
        &3 & 0 & 0
        &35.4\% & 0\%  & 0\%\\
    \hline
    \multicolumn{20}{l}{Note: RQ1, RQ2B, and RQ3B present total correct predictions out of 8, 2, and 2 total samples, respectively.}\\
  \end{tabular}
  }
\end{table}
\begin{table}[H]
\centering
  \caption{Model Performance on RQ1, RQ2B and RQ3B - CWE-352}
  \label{Appendices:RQ1-RQ2B-CWE-352}
  \resizebox{\columnwidth}{!}{
  \begin{tabular}{|l|r|r|r|r|r|r|r|r|r|r|r|r|r|r|r|r|r|r|r|r|r|}
    \hline
    Model & \multicolumn{3}{c|}{Seed 26312} & \multicolumn{3}{c|}{Seed 43511} & \multicolumn{3}{c|}{Seed 67732} & \multicolumn{3}{c|}{Seed 70757} & \multicolumn{3}{c|}{Seed 95541} & \multicolumn{3}{c|}{Seed 123456} & \multicolumn{3}{c|}{Average PP}\\
    \cline{2-22}
    & {\tiny RQ1} & {\tiny RQ2B} & {\tiny RQ3B}
    & {\tiny RQ1} & {\tiny RQ2B} & {\tiny RQ3B}
    & {\tiny RQ1} & {\tiny RQ2B} & {\tiny RQ3B}
    & {\tiny RQ1} & {\tiny RQ2B} & {\tiny RQ3B}
    & {\tiny RQ1} & {\tiny RQ2B} & {\tiny RQ3B}
    & {\tiny RQ1} & {\tiny RQ2B} & {\tiny RQ3B} 
    & {\tiny RQ1} & {\tiny RQ2B} & {\tiny RQ3B}\\
    \hline
    VR    
        &0 & 0 & 0
        &0 & 0 & 0
        &0 & 0 & 0
        &0 & 0 & 0
        &0 & 0 & 0
        &0 & 0 & 0
        &0\% & 0\%& 0\%\\
        \hline
    CB  
        &0 & 0 & 0
        &0 & 0 & 0
        &0 & 0 & 0
        &0 & 0 & 0
        &0 & 0 & 0
        &0 & 0 & 0
        &0\% & 0\%& 0\%\\
        \hline
    GDB 
        &0 & 0 & 0
        &0 & 0 & 0
        &0 & 0 & 0
        &0 & 0 & 0
        &0 & 0 & 0
        &0 & 0 & 0
        &0\%  & 0\%& 0\%\\
        \hline
    UX  
        &0 & 0 & 0
        &0 & 0 & 0
        &0 & 0 & 0
        &0 & 0 & 0
        &0 & 0 & 0
        &0 & 0 & 0
        &0\%  & 0\% & 0\%\\
    \hline
    \multicolumn{21}{l}{Note: RQ1, RQ2B, and RQ3B present total correct predictions out of 2, 2, and 2 total samples, respectively.}\\
  \end{tabular}
  }
\end{table}
\begin{table}[H]
\centering
  \caption{Performance Of Models on RQ5 using RQ3B dataset - Pre-training}
  \label{Appendices:RQ5-pretrain}
  \begin{tabular}{|l|r|r|r|r|}
    \hline
    Model          & Beam = 1 & Beam = 3 & Beam = 5 & Beam = 50\\
    \hline
    VulRepair       & 3.57\% (0.0357)  & 7.29\% (0.0729) & 7.59\% (0.0759)  & 6.55\% (0.0655) \\
    \hline
    CodeBERT        & 2.98\% (0.0298)  & 4.61\% (0.0461) & 5.36\% (0.0536)  & 11.76\% (0.1176) \\
    \hline
    GraphCodeBERT   & 2.23\% (0.0223)  & 4.61\% (0.0461) & 5.8\% (0.058)    & 11.76\% (0.1176) \\
    \hline
    UniXcoder       & 1.93\% (0.0193)  & 5.21\% (0.0521) & 6.55\% (0.0655)  & 11.31\% (0.1131) \\
    \hline
    \multicolumn{5}{l}{Note: Pre-training is only done once on seed 26312.}\\
  \end{tabular}
\end{table}
\begin{table}[H]
\centering
  \caption{VulRepair Performance on RQ5 using RQ3B dataset - Transfer Learning}
  \label{Appendices:RQ5-finetrain-VulRepair}
  \begin{tabular}{|l|r|r|r|r|r|r|r|r|}
    \hline
    Seed          & Beam = 1               & Beam = 3           & Beam = 5            & Beam = 10   & Beam = 20     & Beam = 30     & Beam = 40     & Beam = 50\\
    \hline
    26312          & 0.1324                 & 0.1890             & 0.1964             & 0.2024      & 0.1905        & 0.1845        & 0.1845        & 0.1845 \\
    \hline
    43511          & 0.1354                 & 0.1815             & 0.1875             & 0.1935      & 0.186         & 0.1875        & 0.1801        & 0.1786 \\
    \hline
    67732          & 0.1399                 & 0.1890             & 0.2068             & 0.2024      & 0.1994        & 0.2009        & 0.1994        & 0.1964 \\
    \hline
    70757          & 0.1399                 & 0.1994             & 0.2143             & 0.2068      & 0.2054        & 0.1994        & 0.2024        & 0.2009 \\
    \hline
    95541          & 0.1414                 & 0.1994             & 0.2098             & 0.2128      & 0.2054        & 0.1964        & 0.1905        & 0.186 \\
    \hline
    123456         & 0.1220                 & 0.1830             & 0.1979             & 0.186       & 0.183         & 0.1771        & 0.1786        & 0.1741 \\
    \hline
    Average PP     & 13.53\%                & 19.03\%            & 20.21\%            & 20.0\%      & 19.5\%        & 19.17\%       & 19.0\%        & 18.67\%  \\
    \hline
  \end{tabular}
\end{table}
\begin{table}[H]
\centering
  \caption{CodeBERT Performance on RQ5 using RQ3B dataset - Transfer Learning}
  \label{Appendices:RQ5-finetrain-CodeBERT}
  \begin{tabular}{|l|r|r|r|r|r|r|r|r|}
    \hline
    Seed          & Beam = 1            & Beam = 3          & Beam = 5          & Beam = 10   & Beam = 20     & Beam = 30     & Beam = 40     & Beam = 50\\
    \hline
    26312          & 0.119              & 0.1711            & 0.189             & 0.2068      & 0.2262        & 0.2336        & 0.2411        & 0.2455 \\
    \hline
    43511          & 0.124              & 0.1696            & 0.189             & 0.2054      & 0.2232        & 0.2366        & 0.2411        & 0.2411 \\
    \hline
    67732          & 0.128              & 0.1815            & 0.1935            & 0.2158      & 0.2292        & 0.2351        & 0.2396        & 0.2455 \\
    \hline
    70757          & 0.131              & 0.1756            & 0.2009            & 0.2232      & 0.2396        & 0.244         & 0.247         & 0.253 \\
    \hline
    95541          & 0.128              & 0.1682            & 0.1786            & 0.2039      & 0.2307        & 0.2426        & 0.2455        & 0.2485 \\
    \hline
    123456         & 0.118              & 0.170             & 0.183             & 0.1979      & 0.2143        & 0.2277        & 0.2351        & 0.2396 \\
    \hline
    Average PP     & 12.47\%            & 17.27\%           & 18.9\%            & 20.83\%     & 22.67\%       & 23.67\%       & 24.17\%       & 24.55\%  \\
    \hline
  \end{tabular}
\end{table}
\begin{table}[H]
\centering
  \caption{GraphCodeBERT Performance on RQ5 using RQ3B dataset - Transfer Learning}
  \label{Appendices:RQ5-finetrain-GraphCodeBERT}
  \begin{tabular}{|l|r|r|r|r|r|r|r|r|}
    \hline
    Seed          & Beam = 1            & Beam = 3          & Beam = 5          &  Beam = 10        & Beam = 20         & Beam = 30         & Beam = 40                & Beam = 50\\
    \hline
    26312          & 0.125              & 0.1786            & 0.1949            & 0.2173            & 0.2396            & 0.2485            & 0.2589                   & 0.2619 \\
    \hline
    43511          & 0.128              & 0.1845            & 0.2054            & 0.2247            & 0.2426            & 0.253             & 0.2574                   & 0.2634 \\
    \hline
    67732          & 0.1131             & 0.1756            & 0.1905            & 0.2143            & 0.2351            & 0.244             & 0.253                    & 0.2649 \\
    \hline
    70757          & 0.1131             & 0.1741            & 0.1964            & 0.2188            & 0.2321            & 0.247             & 0.253                    & 0.2574 \\
    \hline
    95541          & 0.1161             & 0.1696            & 0.1905            & 0.2173            & 0.2292            & 0.2396            & 0.2411                   & 0.247  \\
    \hline
    123456         & 0.0952             & 0.131             & 0.1592            & 0.183             & 0.2068            & 0.2143            & 0.2202                   & 0.2307 \\
    \hline
    Average PP     & 11.51\%            & 16.89\%           & 18.95\%           & 21.33\%          & 23.17\%            & 24.17\%           & 24.67\%                  & 25.42\% \\
    \hline
  \end{tabular}
\end{table}
\begin{table}[H]
\centering
  \caption{UniXcoder Performance on RQ5 using RQ3B dataset - Transfer Learning}
  \label{Appendices:RQ5-finetrain-UniXcoder}
  \begin{tabular}{|l|r|r|r|r|r|r|r|r|}
    \hline
    Seed          & Beam = 1          & Beam = 3           & Beam = 5          & Beam = 10   & Beam = 20     & Beam = 30     & Beam = 40              & Beam = 50\\
    \hline
    26312          & 0.1339            & 0.1741             & 0.1964            & 0.2202     & 0.2381        & 0.244         & 0.25                   & 0.256 \\
    \hline
    43511          & 0.1369            & 0.189              & 0.2024            & 0.2262     & 0.247         & 0.2515        & 0.2604                 & 0.2649 \\
    \hline
    67732          & 0.1339            & 0.1875             & 0.2054            & 0.2307     & 0.2426        & 0.253         & 0.2649                 & 0.2723 \\
    \hline
    70757          & 0.1250            & 0.1741             & 0.1949            & 0.2232     & 0.2426        & 0.256         & 0.2619                 & 0.2664 \\
    \hline
    95541          & 0.1235            & 0.1845             & 0.1964            & 0.2202     & 0.2381        & 0.247         & 0.253                  & 0.256 \\
    \hline
    123456         & 0.1235            & 0.1786             & 0.1890            & 0.2128     & 0.2307        & 0.2381        & 0.2455                 & 0.2485 \\
    \hline
    Average PP     & 12.94\%           & 18.13\%            & 19.74\%           & 22.17\%    & 24.0\%        & 24.83\%       & 25.67\%                & 26.07\% \\
    \hline
  \end{tabular}
\end{table}

\twocolumn

\bibliographystyle{IEEEtran}
\bibliography{ACSAC2024_bibliography}

\begin{thebibliography}{10}
\providecommand{\url}[1]{#1}
\csname url@samestyle\endcsname
\providecommand{\newblock}{\relax}
\providecommand{\bibinfo}[2]{#2}
\providecommand{\BIBentrySTDinterwordspacing}{\spaceskip=0pt\relax}
\providecommand{\BIBentryALTinterwordstretchfactor}{4}
\providecommand{\BIBentryALTinterwordspacing}{\spaceskip=\fontdimen2\font plus
\BIBentryALTinterwordstretchfactor\fontdimen3\font minus \fontdimen4\font\relax}
\providecommand{\BIBforeignlanguage}[2]{{%
\expandafter\ifx\csname l@#1\endcsname\relax
\typeout{** WARNING: IEEEtran.bst: No hyphenation pattern has been}%
\typeout{** loaded for the language `#1'. Using the pattern for}%
\typeout{** the default language instead.}%
\else
\language=\csname l@#1\endcsname
\fi
#2}}
\providecommand{\BIBdecl}{\relax}
\BIBdecl

\bibitem{chen2022neural}
Z.~Chen, S.~Kommrusch, and M.~Monperrus, ``Neural transfer learning for repairing security vulnerabilities in c code,'' \emph{IEEE Transactions on Software Engineering}, vol.~49, no.~1, pp. 147--165, 2022.

\bibitem{fu2022vulrepair}
M.~Fu, C.~Tantithamthavorn, T.~Le, V.~Nguyen, and D.~Phung, ``Vulrepair: a t5-based automated software vulnerability repair,'' in \emph{Proceedings of the 30th ACM joint european software engineering conference and symposium on the foundations of software engineering}, 2022, pp. 935--947.

\bibitem{zhang2023pre}
Q.~Zhang, C.~Fang, B.~Yu, W.~Sun, T.~Zhang, and Z.~Chen, ``Pre-trained model-based automated software vulnerability repair: How far are we?'' \emph{IEEE Transactions on Dependable and Secure Computing}, 2023.

\bibitem{huang2023empirical}
K.~Huang, X.~Meng, J.~Zhang, Y.~Liu, W.~Wang, S.~Li, and Y.~Zhang, ``An empirical study on fine-tuning large language models of code for automated program repair,'' in \emph{2023 38th IEEE/ACM International Conference on Automated Software Engineering (ASE)}.\hskip 1em plus 0.5em minus 0.4em\relax IEEE, 2023, pp. 1162--1174.

\bibitem{fu2024vision}
M.~Fu, V.~Nguyen, C.~Tantithamthavorn, D.~Phung, and T.~Le, ``Vision transformer inspired automated vulnerability repair,'' \emph{ACM Transactions on Software Engineering and Methodology}, vol.~33, no.~3, pp. 1--29, 2024.

\bibitem{mastropaolo2024training}
A.~Mastropaolo, V.~Nardone, G.~Bavota, and M.~Di~Penta, ``How the training procedure impacts the performance of deep learning-based vulnerability patching,'' \emph{arXiv preprint arXiv:2404.17896}, 2024.

\bibitem{ding2024vulnerability}
Y.~Ding, Y.~Fu, O.~Ibrahim, C.~Sitawarin, X.~Chen, B.~Alomair, D.~Wagner, B.~Ray, and Y.~Chen, ``Vulnerability detection with code language models: How far are we?'' \emph{arXiv preprint arXiv:2403.18624}, 2024.

\bibitem{fan2020ac}
J.~Fan, Y.~Li, S.~Wang, and T.~N. Nguyen, ``Ac/c++ code vulnerability dataset with code changes and cve summaries,'' in \emph{Proceedings of the 17th International Conference on Mining Software Repositories}, 2020, pp. 508--512.

\bibitem{bhandari2021cvefixes}
G.~Bhandari, A.~Naseer, and L.~Moonen, ``Cvefixes: automated collection of vulnerabilities and their fixes from open-source software,'' in \emph{Proceedings of the 17th International Conference on Predictive Models and Data Analytics in Software Engineering}, 2021, pp. 30--39.

\bibitem{vqm2024}
\BIBentryALTinterwordspacing
M.~Fu. (2024) {VQM}. [Online]. Available: \url{https://github.com/awsm-research/VQM}
\BIBentrySTDinterwordspacing

\bibitem{croft2023data}
R.~Croft, M.~A. Babar, and M.~M. Kholoosi, ``Data quality for software vulnerability datasets,'' in \emph{2023 IEEE/ACM 45th International Conference on Software Engineering (ICSE)}.\hskip 1em plus 0.5em minus 0.4em\relax IEEE, 2023, pp. 121--133.

\bibitem{iso/iec25012}
{ISO/IEC}, ``{Software engineering--Software product Quality Requirements and Evaluation (SQuaRE)--Data quality models},'' 2008.

\bibitem{acm-badges}
\BIBentryALTinterwordspacing
{ACM}. (2020) Artifact review and badging - current. [Online]. Available: \url{https://www.acm.org/publications/policies/artifact-review-and-badging-current}
\BIBentrySTDinterwordspacing

\bibitem{VulRepair-HuggingFace}
\BIBentryALTinterwordspacing
M.~Fu. (2022) {VulRepair}. [Online]. Available: \url{https://huggingface.co/datasets/MickyMike/cvefixes_bigvul}
\BIBentrySTDinterwordspacing

\bibitem{wang2021codet5}
Y.~Wang, W.~Wang, S.~Joty, and S.~C. Hoi, ``Codet5: Identifier-aware unified pre-trained encoder-decoder models for code understanding and generation,'' \emph{arXiv preprint arXiv:2109.00859}, 2021.

\bibitem{feng2020codebert}
Z.~Feng, D.~Guo, D.~Tang, N.~Duan, X.~Feng, M.~Gong, L.~Shou, B.~Qin, T.~Liu, D.~Jiang \emph{et~al.}, ``Codebert: A pre-trained model for programming and natural languages,'' \emph{arXiv preprint arXiv:2002.08155}, 2020.

\bibitem{guo2020graphcodebert}
D.~Guo, S.~Ren, S.~Lu, Z.~Feng, D.~Tang, S.~Liu, L.~Zhou, N.~Duan, A.~Svyatkovskiy, S.~Fu \emph{et~al.}, ``Graphcodebert: Pre-training code representations with data flow,'' \emph{arXiv preprint arXiv:2009.08366}, 2020.

\bibitem{guo2022unixcoder}
D.~Guo, S.~Lu, N.~Duan, Y.~Wang, M.~Zhou, and J.~Yin, ``Unixcoder: Unified cross-modal pre-training for code representation,'' \emph{arXiv preprint arXiv:2203.03850}, 2022.

\bibitem{top10}
\BIBentryALTinterwordspacing
comparitech. (2021) 25+ cyber security vulnerability statistics and facts. [Online]. Available: \url{25+ cyber security vulnerability statistics and facts}
\BIBentrySTDinterwordspacing

\bibitem{herzig2013impact}
K.~Herzig and A.~Zeller, ``The impact of tangled code changes,'' in \emph{2013 10th Working Conference on Mining Software Repositories (MSR)}.\hskip 1em plus 0.5em minus 0.4em\relax IEEE, 2013, pp. 121--130.

\bibitem{vulrepairgit}
\BIBentryALTinterwordspacing
awsm research. (2022) {VulRepair}. [Online]. Available: \url{https://github.com/awsm-research/VulRepair}
\BIBentrySTDinterwordspacing

\bibitem{isenglabgit}
\BIBentryALTinterwordspacing
{ISEngLab}. (2023) {LLM4VulFix}. [Online]. Available: \url{https://github.com/isenglab/llm4vulfix}
\BIBentrySTDinterwordspacing

\bibitem{chen2023diversevul}
Y.~Chen, Z.~Ding, L.~Alowain, X.~Chen, and D.~Wagner, ``Diversevul: A new vulnerable source code dataset for deep learning based vulnerability detection,'' in \emph{Proceedings of the 26th International Symposium on Research in Attacks, Intrusions and Defenses}, 2023, pp. 654--668.

\end{thebibliography}

\end{document}